\documentclass[twocolumn]{aastex631}

\newcommand{\water}{H$_2$O}

\newcommand{\methane}{CH$_4$}
\newcommand{\sot}{SO$_2$}
\newcommand{\methanol}{CH$_3$OH}

\usepackage{float}
\usepackage{subfigure}
\usepackage{graphicx} 
\usepackage{scalerel}
\usepackage{natbib}
\usepackage{float}
\usepackage{booktabs}
\usepackage{array}
\usepackage{hyperref}
\usepackage{textcomp}  
\usepackage{booktabs}

\begin{document}

\title{CORINOS IV: Quantifying Baseline-Fitting Uncertainties in \sot\ Ice Measurements with JWST/MIRI}

\correspondingauthor{Rachel E. Gross}
\email{reg4ff@virginia.edu}

\author[0000-0002-0477-6047]{Rachel E. Gross}
\affiliation{University of Virginia, Department of Chemistry, Charlottesville, VA 22904, USA}

\author[0000-0001-8227-2816]{Yao-Lun Yang}
\affiliation{Star and Planet Formation Laboratory, RIKEN Pioneering Research Institute, Wako-shi, Saitama, 351-0198, Japan}

\author[0000-0003-2076-8001]{L.Ilsedore Cleeves}
\affiliation{University of Virginia, Department of Astronomy, Charlottesville, VA 22904, USA}
\affiliation{University of Virginia, Department of Chemistry, Charlottesville, VA 22904, USA}

\author[0000-0001-7591-1907]{Ewine F. van Dishoeck}
\affiliation{Leiden Observatory, Leiden University, PO Box 9513, NL 2300 RA Leiden, The Netherlands}
\affiliation{Max Planck Institute for Extraterrestrial Physics, Garching, Germany}

\author[0000-0001-7723-8955]{Robin T. Garrod}
\affiliation{Departments of Chemistry and Astronomy, University of Virginia, Charlottesville, VA 22904, USA}

\author[0000-0002-4801-436X]{Mihwa Jin}
\affiliation{Astrochemistry Laboratory, Code 691, NASA Goddard Space Flight Center, Greenbelt, MD 20771, USA}
\affiliation{Department of Physics, Catholic University of America, Washington, DC 20064, USA}

\author[0000-0002-3297-4497]{Nami Sakai}
\affiliation{Star and Planet Formation Laboratory, RIKEN Pioneering Research Institute, Wako-shi, Saitama, 351-0198, Japan}

\author[0000-0002-5171-7568]{Christopher N. Shingledecker}
\affiliation{Department of Chemistry, Virginia Military Institute, Lexington, VA 24450, USA}

\author[0000-0001-8064-2801]{JaeYeong Kim}
\affiliation{Korea Astronomy and Space Science Institute, 776 Daedeok-daero, Yuseong-gu, Daejeon 34055, Republic of Korea}

\author[0000-0002-8716-0482]{Jennifer B. Bergner}
\affiliation{Department of the Geophysical Sciences, University of Chicago, Chicago, IL 60637, USA}

\author[0000-0001-5175-1777]{Neal J. Evans II}
\affiliation{Department of Astronomy, The University of Texas at Austin, Austin, TX 78712, USA}

\author[0000-0003-1665-5709]{Joel D. Green}
\affiliation{Space Telescope Science Institute, Baltimore, 3700 San Martin Dr., MD 21218, USA}

\author[0000-0002-2523-3762]{Chul-Hwan Kim}
\affiliation{Department of Physics and Astronomy, Seoul National University, 1 Gwanak-ro, Gwanak-gu, Seoul 08826, Korea}

\author[0000-0003-3119-2087]{Jeong-Eun Lee}
\affiliation{Department of Physics and Astronomy, Seoul National University, 1 Gwanak-ro, Gwanak-gu, Seoul 08826, Korea}

\author[0000-0003-3655-5270]{Yuki Okoda}
\affiliation{NRC Herzberg Astronomy and Astrophysics, 5071 West Saanich Road, Victoria, BC, V9E 2E7, Canada}
\affiliation{Star and Planet Formation Laboratory, RIKEN Cluster for Pioneering Research, Wako-shi, Saitama, 351-0198, Japan}

\author[0000-0003-4147-4125]{Will R.M. Rocha}
\affiliation{Laboratory for Astrophysics, Leiden Observatory, Leiden University, PO Box 9513, NL 2300 RA Leiden, The Netherlands}

\author[0000-0003-4147-4125]{Brielle Shope}
\affiliation{University of Virginia, Department of Chemistry, Charlottesville, VA 22904, USA}

\author[0000-0002-9497-8856]{Himanshu Tyagi}
\affiliation{Department of Astronomy and Astrophysics, Tata Institute of Fundamental Research, Homi Bhabha Road, Colaba, Mumbai 400005, India}




\begin{abstract}

Sulfur dioxide (SO$_2$) ice has been tentatively detected in protostellar envelopes, but its reliability as a solid-state sulfur reservoir remains unclear. We present new measurements of SO$_2$ ice from 6.8-8.5~$\mu$m toward four Class 0 protostars observed with JWST's Mid-Infrared (MIRI) Medium Resolution Spectrometer, as part of the COMs ORigin Investigated by the Next-generation Observatory in Space (CORINOS) program. The sample spans a luminosity range from 1~$L_\odot$ (B335, IRAS~15398-3359) to 10~$L_\odot$ (L483, Ser-emb~7). To assess continuum placement uncertainty in absorption spectra, we apply randomized polynomial fits over the restricted region. We fit laboratory spectra from the Leiden Ice Database for Astrochemistry (LIDA) using the open-source Python library \texttt{Omnifit}. We detect the 7.7~$\mu$m CH$_4$ band in all sources and find its column density robust to baseline choice, providing a reference for evaluating the weaker SO$_2$ feature on its blue shoulder and quantifying baseline-related uncertainty. Three SO$_2$ laboratory ices were tested: pure SO$_2$ ice yields 0.3-1.2\% of volatile sulfur may be locked in SO$_2$ ice (lower and upper limits); CH$_3$OH:SO$_2$ ice gives 0.02-0.18\%, but with lower quality fitting. The best-fitting H$_2$O:SO$_2$ ice yields 0.2-0.9\%, which we consider the most realistic. These ranges define plausible bounds on SO$_2$ ice abundances in our sample. We find evidence for SO$_2$ in Ser-emb~7, L483, and IRAS~15398-3359, but emphasize the noisy spectrum of B335 prevents a definitive detection. Comparing SO$_2$ ice abundances across the different environments, we assess how conditions influence role of SO$_2$ as a potential sulfur reservoir and implications for the longstanding ``missing sulfur'' problem.

\end{abstract}



\section{Introduction} \label{intro}

Sulfur-bearing molecules are detected throughout the interstellar medium and across all stages of star formation, from dense clouds \citep[e.g.,][]{Navarro-Almaida2020,Esplugues2022} to protostars \citep[e.g.,][]{Blake1987,Drozdovskaya2018} to protoplanetary disks \citep[e.g.,][]{Fuente2010,Semenov2018,laas2019} and comets \citep[e.g.,][]{Smith1980,Altwegg2022}. Despite sulfur’s cosmic abundance (S/H $\sim$1.3$\times$10$^{-5}$), observations in dense environments consistently recover only a small fraction of the expected sulfur budget \citep{tieftrunk1994,wakelam2004,Anderson2013,mcclure2023ices}. This long-standing depletion has motivated the idea that most sulfur resides in solid material on or within interstellar ices, yet its specific carriers remain uncertain. To date, OCS is the only sulfur-bearing species securely identified in the solid phase \citep[e.g.,][]{Palumbo1995,Boogert2015,mcclure2023ices, Boogert2022,Santos2024b,Taillard2025}. A weak absorption feature near 7.6~$\mu$m, first attributed to SO$_2$ in W33A and NGC~7538 IRS~1 \citep{Boogert1997}, has since been tentatively reported toward both molecular clouds \citep{mcclure2023ices} and protostars \citep{rocha2024}, but the SO$_2$ band remains difficult to isolate because of blending with CH$_4$, OCN$^-$, and complex organics. As a result, its underlying contribution is still uncertain, and further quantifying the SO$_2$ ice feature and its sensitivity to local protostellar condition will provide constraints for evolving astrochemical models to better predict how sulfur is distributed during planet formation.

Interpreting a weak feature like \sot\ can be challenging due to both overlap with other molecular ices and needing accurate estimates of the ``ice-free'' baseline \citep[e.g.,][]{gibb2004,Schutte1999,boogert2008}. On the first point, the SO$_2$ stretching mode near 7.6~$\mu$m is weak and blended with absorption from CH$_4$, OCN$^-$, and complex organics, making it particularly sensitive to analysis \citep[e.g.,][]{rocha2024}. On the second point, all ice absorption measurements require some estimate of the absorption-free continuum to accurately estimate ice optical depth and thus ice content. Protostellar ice features particularly in the mid-infrared are often broad and overlapping, making identification of relatively clean bandwidth challenging. As a result, small changes in baseline placement can significantly affect the shapes, depths, and integrated areas of absorption bands, leading to uncertainties in derived abundances \citep{Boogert1997,rocha2024}. The spectral window between 6.8–8.6~$\mu$m is especially crowded, containing the CH$_4$ deformation mode at 7.7~$\mu$m \citep{oberg2008} along with deformation and bending modes of ethanol (CH$_3$CH$_2$OH), formate (HCOO$^-$), and acetaldehyde (CH$_3$CHO) at 7.24 and 7.43~$\mu$m \citep{Schutte1999,TvS2018}. As a result, the SO$_2$ shoulder at 7.63~$\mu$m is susceptible to blending with multiple species, making robust identification difficult with previous facilities.

Despite these challenges, observational studies with earlier infrared telescopes, such as the Infrared Space Observatory (ISO; \citep{gibb2004}) and Spitzer IRS \citep{boogert2008,Boogert2015}, have provided the data for much of our known ice inventory, including major mid-IR absorbers such as \water, CO, CO$_2$, \methane, and NH$_3$. Ground-based facilities like ESO's Very Large Telescope (VLT), Subaru, Keck, and NASA's Infrared Telescope Facility (IRTF), as well as AKARI, have further advanced our understanding of interstellar ices through targeted surveys, including large VLT programs of CO ice \citep{pontoppidan2003} and OCN$^-$ ice \citep{vanBroekhuizen2005}, along with broader efforts spanning multiple species \citep{grim1991,dartois1999a,Dartois1999b, shimonishi2010,Chu2020,perotti2021}. However, the sensitivity and resolving power of these facilities were insufficient to reliably isolate SO$_2$ from the CH$_4$ feature.

The improved sensitivity and spectral resolution of the James Webb Space Telescope (JWST) has revolutionized the study of ices. JWST programs have already begun to transform our knowledge of ices. The IceAge Early Release Science program (PID 1128) has provided initial ice compositions in dense molecular clouds before protostellar formation \citep{mcclure2023ices}. The Guaranteed Time Observations Program JOYS \citep[PID 1290;][]{rocha2024,Chen2024,vanDishoeck2025} targeted high and low-mass Class 0/I protostars. \citet{rocha2024} applied laboratory data fitting with the ENIIGMA genetic-algorithm tool \citep{Rocha_eniigma}, while \citet{Chen2024} analyzed the same JOYS data using a two-step spectral-fitting routine that combined least-squares and Markov chain Monte Carlo (MCMC) methods \citep[e.g.,][]{Foreman-Mackey2013}. These advances now allow weak and blended mid-infrared ice features to be measured with greater accuracy than was possible with previous facilities.

Complex organic molecules (COMs), which are carbon-bearing species containing at least six atoms, are a link between simple ice species and the prebiotic building blocks that allow life as we know it to occur. COMs form efficiently via reactions of simple ices on dust grains, and then sublimate via numerous pathways to be detected in the gas \citep[e.g.,][]{herbst2009,jorgensen2016,Jorgensen2020,vandishoeck2021,herbst_vandishoeck_planetformation2021}. COM detection in the solid phase, along with the simple species from which they form, has been challenging due to broad and overlapping vibrational modes in the infrared. Among COMs, methanol (CH$_3$OH) is the simplest COM that has been securely identified in interstellar ices across multiple vibrational bands \citep{grim1991,Dartois2003,pontoppidan2004,oberg2011_spitzer}. Larger COMs such as ethanol (CH$_3$CH$_2$OH) and acetaldehyde (CH$_3$CHO) have been more difficult to identify due to their weaker absorption features and their overlap with stronger modes from abundant simple ices \citep{Schutte1999, yang2022,mcclure2023ices}. Despite extensive laboratory study identifying multiple features, observational detections have remained tentative and rely on single-feature absorption \cite[e.g.,][]{TvS2018}. In the context of this study, these COM bands are important primarily because they contribute to the blended absorption around 7.6–7.7~$\mu$m that complicates SO$_2$ identification.

In this study, we present a new method to evaluate the effect of baseline uncertainty on column density estimates, focusing on the \sot\ feature near 7.63~$\mu$m. We apply this approach to JWST MIRI data from the CORINOS Cycle 1 General Observer program \citep{yang2022}, which targets four isolated Class 0 protostars that span a range of luminosities and gas-phase COM activity. By conducting multiple baseline fits and modeling the resulting spectra with laboratory ice data, we assess the robustness of \sot\ detections and place constraints on its abundance relative to \methane. The paper is organized as follows: Section \ref{sources} introduces the target sources; Section \ref{observations} describes the data reduction process with further details in \citet{yang2022}; Section \ref{methods} outlines our baseline fitting and modeling approach; Section \ref{results} presents the resulting fits and column density estimates; and Section \ref{discussion} explores the implications for the missing sulfur problem and interpreting our results.

\section{Sources} \label{sources}
\subsection{IRAS 15398-3359}

IRAS 15398-3359 (B228) is a Class 0 protostar located in the Lupus I molecular cloud \citep[e.g.,][]{Heyer1989} at a distance of 154.9 $\pm$ 3.4 pc \citep{Galli2020}. With a bolometric luminosity of 1.4 $L_\odot$ and bolometric temperature of $T_\mathrm{bol}=50$~K \citep{Ohashi2023}, with an envelope mass of 1.2~$M_\odot$ \citep{Jorgensen2013}. IRAS 15398-3359 serves as a key target for understanding early-stage low-mass star formation. Observations have detected molecular line emission in the outflows of this source \citep[e.g.,][]{Okoda2021,Thieme2023}. Simple ices like \water, \methane, and \methanol\ have also been detected with Spitzer in the envelope of this source \citep[e.g.,][]{oberg2008,boogert2008}.
IRAS 15398-3359 has a rich inventory of warm carbon-chain molecules (CCMs), indicative of active warm carbon-chain chemistry (WCCC) \citep{Sakai2009}. This chemical behavior is consistent with models in which CH$_4$ ice formed in the prestellar phase is desorbed during protostellar heating, driving the formation of carbon-chain species \citep{Aikawa2008, Sakai2008}. Episodic accretion events may further contribute to this process by temporarily increasing the temperature in the envelope \citep{Jorgensen2013}, possibly enhancing thermal desorption of ice species.

\subsection{Ser-emb 7}

Ser-emb 7 is a Class 0 protostar located in the Serpens Main star-forming region at a distance of $\sim436$ pc \citep{Ortiz-Leon2017}. It has a relatively high bolometric luminosity of 7.9 $L_\odot$ \citep{Enoch2011}, bolometric temperature of 58 K \citep{Enoch2009} and calculated as a low-inclination system $\sim 25$ degrees. Ser-emb 7 has an envelope mass of $\sim4.3$~$M_\odot$ \citep{Enoch2011}. This source exhibits evidence of compact continuum emission and outflow activity, though detailed constraints on the inner envelope and disk structure remain limited \citep{Enoch2011}. To date, there are no sulfur rich species in either gas or ice detected in this source \citep{Bergner2019}.

\subsection{L483}

L483 is a Class 0/I protostar located in the Aquila Rift at a distance of $\sim 200$ pc \citep{Ortiz-Leon2017,Jacobsen2019}. It has a bolometric luminosity of $\sim 13$~$L_\odot$ and a bolometric temperature of $T_\mathrm{bol}\sim 50$--70 K \citep{Jorgensen2002, Lee2015}, with an envelope mass of 4.4~$M_\odot$ \citep{jorgensen2004}. L483 is embedded in a dense, asymmetric envelope and drives a prominent bipolar outflow seen in CO and H$_2$ line emission \citep{Tafalla2000}. It also exhibits strong ice absorption features and rich COM chemistry in both the gas \citep{Jacobsen2019} and solid phase \citep{Chu2020}. Recent ALMA observations have revealed that L483 is in fact a \emph{binary system} with a projected separation of $\sim 34$ au \citep{Cox2022,Hirota2025}. No circumbinary disk larger than $\sim 30$ au has been detected \citep{Jacobsen2019,Cox2022}, and it remains unclear which component hosts the compact Keplerian disk previously inferred from continuum and C$^{18}$O kinematics \citep{Jacobsen2019}. The system shows complex velocity structure, including SiO emission peaks located $\sim 100$ au northeast (blue-shifted) and $\sim 200$ au north (near systemic velocity) of the continuum peak, with their driving source still under debate \citep{Hirota2025}. Nevertheless, the overall north-blue / south-red velocity gradient traced by C$^{18}$O, SO, and CS remains consistent with a moderately inclined rotating envelope. Observations with the IRAM 30m telescope found several sulfur-bearing species in the gas phase including CH$_3$SH, HCS, HSC, C$_2$S, H$_2$CS, and NS$^+$, with CS being the most abundant \citep{Agundez2019}. SO, OCS, and SO$_2$ were also detected at abundances comparable to those observed in cold pre-stellar cores such as TMC-1 \citep{Agundez2019}.

\subsection{B335} 

B335 is a nearby isolated Class 0 protostar at a distance of $\sim 100$ pc with a low bolometric luminosity of $\sim 1.5$ $L_\odot$ and is nearly edge-on with an inclination of $\sim $87 degrees \citep{Evans2015, Yen2015}, and protostellar mass of 0.25~$M_\odot$ and a disk mass of 0.063~$M_\odot$ \citep{Evans2023}, with an envelope mass of 2.4~$M_\odot$ \citep{Saito1999}. Its bolometric temperature is $\sim 37$ K \citep{Okoda2024}. B335 has been studied as a prototypical example of inside-out collapse and drives a slow, wide-angle bipolar outflow \citep{Zhou1993,Harvey2001}. B335 shows rich organic and deuterated chemistry \citep{Imai2016,Okoda2022_B335}. It exhibits strong water and CO$_2$ ice absorption bands and faint features from methanol and ammonia \citep{ boogert2008}. Gas-phase \sot\ has been detected within 15~au of the protostar with ALMA \citep{Bjerkeli2019}. B335 is an important low-mass source for studying ice chemistry in isolated environments, and, as shown by \citet{Lee2025_B335}, its decade-long luminosity burst provides a unique opportunity to investigate how complex organic molecules and ices respond to protostellar variability.

\begin{figure*}
  \centering
  \includegraphics[scale=0.53]{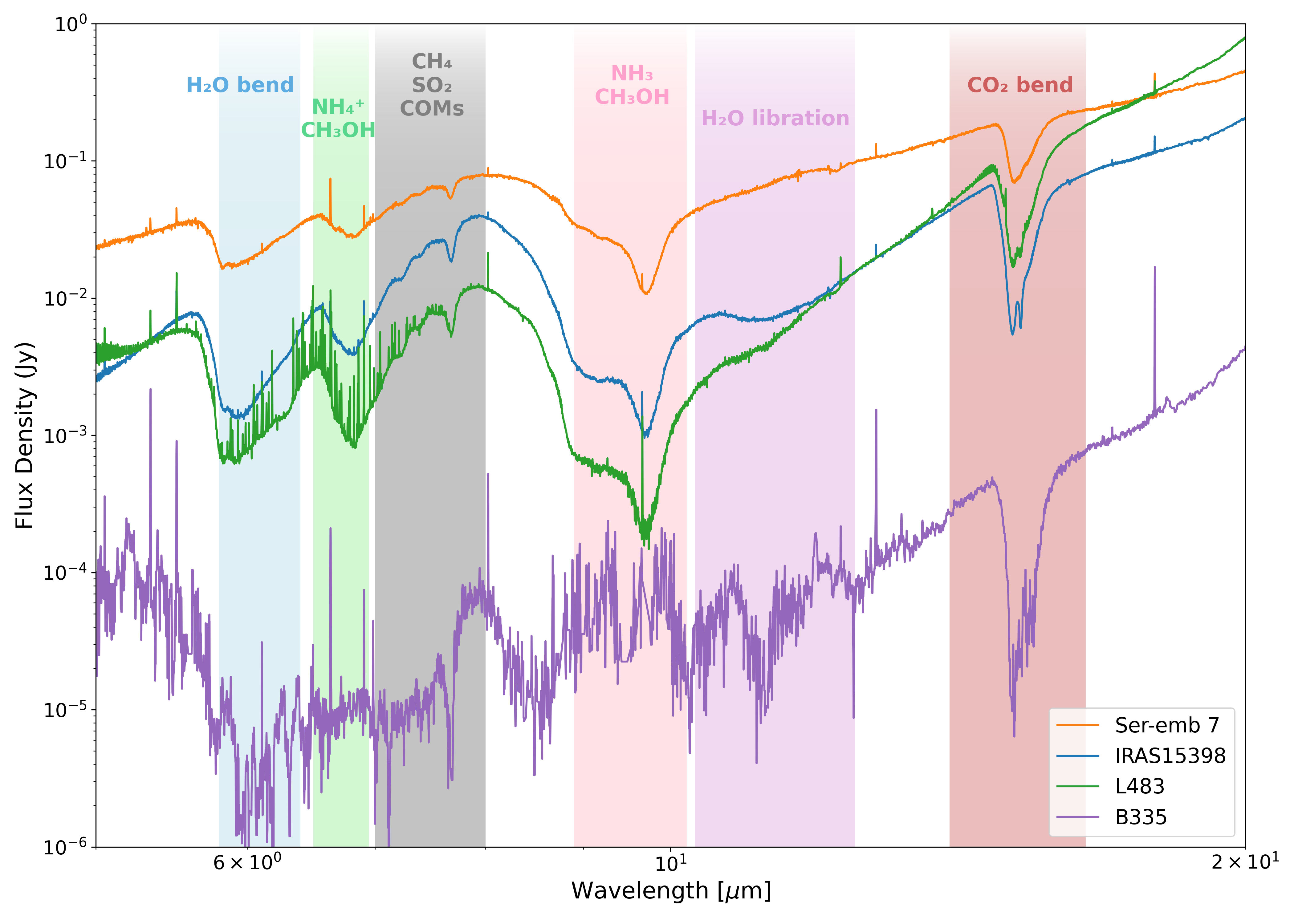}
  \caption{The flux density for all four sources in the CORINOS program. The grey highlighted region between 6.8-8.5 \textmu{}m shows the blended COM region that this paper will focus. These are all 1D spectra extracted with a four-beam aperture with scaling for three of the sources (IRAS 15398-3359, Ser-emb 7, and L483). The sharp peaks in the spectra are from gas-phase species. Note the faintest source, B335, has no subband to subband scaling, and the spectrum is shown as extracted from individual MRS subbands.} 
  \label{fluxdensity_all_labeled}
\end{figure*}

\section{Observations} \label{observations}
All data presented in this paper were taken as part of program ID 2151 with the Mid-InfraRed Instrument (MIRI) \citep[e.g.,][]{Rieke2015,Wright2015} in medium resolution mode (MRS) \citep[e.g.,][]{Wells2015}.
For this program, all four integral field units (IFU) of the MIRI Medium Resolution Spectrometer (MRS) were used, covering 4.9 to 28~$\mu$m. The observations allocated 1433.4 seconds to both the SHORT and LONG wavelength ranges, and 3631.3 seconds to the MEDIUM range to allow higher signal/noise. The data were taken with the SLOWR1 standard read-out mode with four-point dither pattern \citep[see also][for further details]{yang2022}. Some or all of the data presented in this article were obtained from the Mikulski Archive for Space Telescopes (MAST) at the Space Telescope Science Institute. The specific observations analyzed can be accessed via \dataset[DOI: 10.17909/77j7-pd29]{ https://doi.org/10.17909/77j7-pd29}. Data reduction was conducted using the JWST calibration pipeline version 1.15.1 \citep{Bushouse2025}, following methodologies outlined in \citet{yang2022}. The process included pixel-to-pixel background subtraction and fringe reduction steps in Stage 2. The outcome comprised 12 spectral cubes across the three wavelength ranges. Spectrum extraction was performed using \texttt{photutils} \citep{Bradley2023}, adopting a method that employs an aperture size four times the beam diameter, ranging from 0.38'' to 2.17''. For B335, which is significantly fainter and noisier, we used a fixed 1'' aperture and additionally tested an extraction from a 1'' offset position to reduce background contamination. The extracted spectra were then normalized by scaling flux densities from the shortest to the longest wavelengths to align median values across overlapping spectral segments. For B335 we did not apply subband scaling. The higher noise in each spectrum, which becomes worse toward the edges of each subband, makes the median-matching method unreliable and the scaling performs poorly. Without any scaling we do not see a substantial offset between subbands, suggesting that the original flux agreement is acceptable. The different appearance of the B335 spectrum is therefore mainly due to its higher noise level rather than intrinsic spectral structure. The resulting spectra typically reach rms noise levels of $\sim$0.1\,mJy in the 5–15~$\mu$m region, increasing to $\sim$0.15–0.2\,mJy toward the longest wavelengths. These values are consistent with previous MIRI ice studies at similar integration times \citep[e.g.,][]{rocha2024}.

\section{Methodology} \label{methods}

\subsection{Baseline fitting}\label{baseline}

The first step in analyzing ice absorption spectra is to obtain an optical depth spectrum and isolate absorption features from the underlying background. Once the continuum is fit, then the optical depth is calculated using the standard relation:
\begin{equation}
\tau(\lambda) = -\ln\left(\frac{F(\lambda)}{C(\lambda)}\right)
\label{opticaldepth}
\end{equation}
\noindent where $F(\lambda)$ is the observed flux density and $C(\lambda)$ is the fitted continuum. 

Figure \ref{fluxdensity_all_labeled} illustrates the challenge of identifying a ``clean'' baseline in ice rich sources. All four sources increase in flux density with wavelength across the MIRI MRS range, consistent with embedded protostars. The strongly absorbed features observed in the spectra are attributed to various ice species including a broad water bending mode near 6~$\mu$m and silicate absorption near 10~$\mu$m.  Between 7 and 8~$\mu$m lies the COM fingerprint region, which is the focus of our study, containing many species such as methane (\methane), sulfur dioxide (\sot), formic acid (HCOOH), ammonium (NH$_4$$^+$), acetic acid (CH$_3$COOH), and acetaldehyde (CH$_3$CHO), among others.

In ice rich regions like those described here, two approaches are taken, either a global approach that typically involves fitting a guided polynomial across the entire MIRI spectrum \citep[e.g.,][]{boogert2008,mcclure2023ices,Rocha2025} or alternatively a local approach that breaks the spectrum into manageable ranges to better isolate the features of interest, particularly useful for blended features \citep{Schutte1999,rocha2024}.
\citet{Rayalacheruvu2025} introduced INDRA, a Python-based global fitting tool that performs continuum and silicate removal, then applies weighted non-negative least squares to fit the full MIRI spectrum, providing column densities and significance estimates while extending ice inventories to include organic refractories.

For the present study, we have opted to take a local approach since the features of interest \sot\ and \methane, are relatively weak and global fits can become dominated by fits to the strongest features. We specifically narrow our focus to the 6.8-8.5~$\mu$m spectral region and 
do not explicitly fit the silicate feature, as it primarily contributes to absorption beyond 9~$\mu$m.  We first pre-process the data by removing emission lines using a median filter \citep{huang1979} implemented with SciPy’s \texttt{median\_filter} routine \citep{Virtanen2020}. In a median filter, the kernel size specifies the number of data points over which the median is computed. The filtering parameters were customized for each source: IRAS 15398 was smoothed using a 5-point kernel; for B335, NaN values were first removed, the sharp emission peaks masked and clipped followed by a median filter with a 5-point kernel; L483, which also contains prominent atomic and molecular emission lines, was first smoothed with a Savitzky-Golay filter \citep{Savitzky1964} as implemented in SciPy’s \texttt{savgol\_filter} routine \citep{Virtanen2020}, then processed with a median filter with an 11-point kernel; and Ser-emb 7 was smoothed using a 5-point kernel.

\begin{figure*}[t]
  \centering
  \includegraphics[scale=0.6]{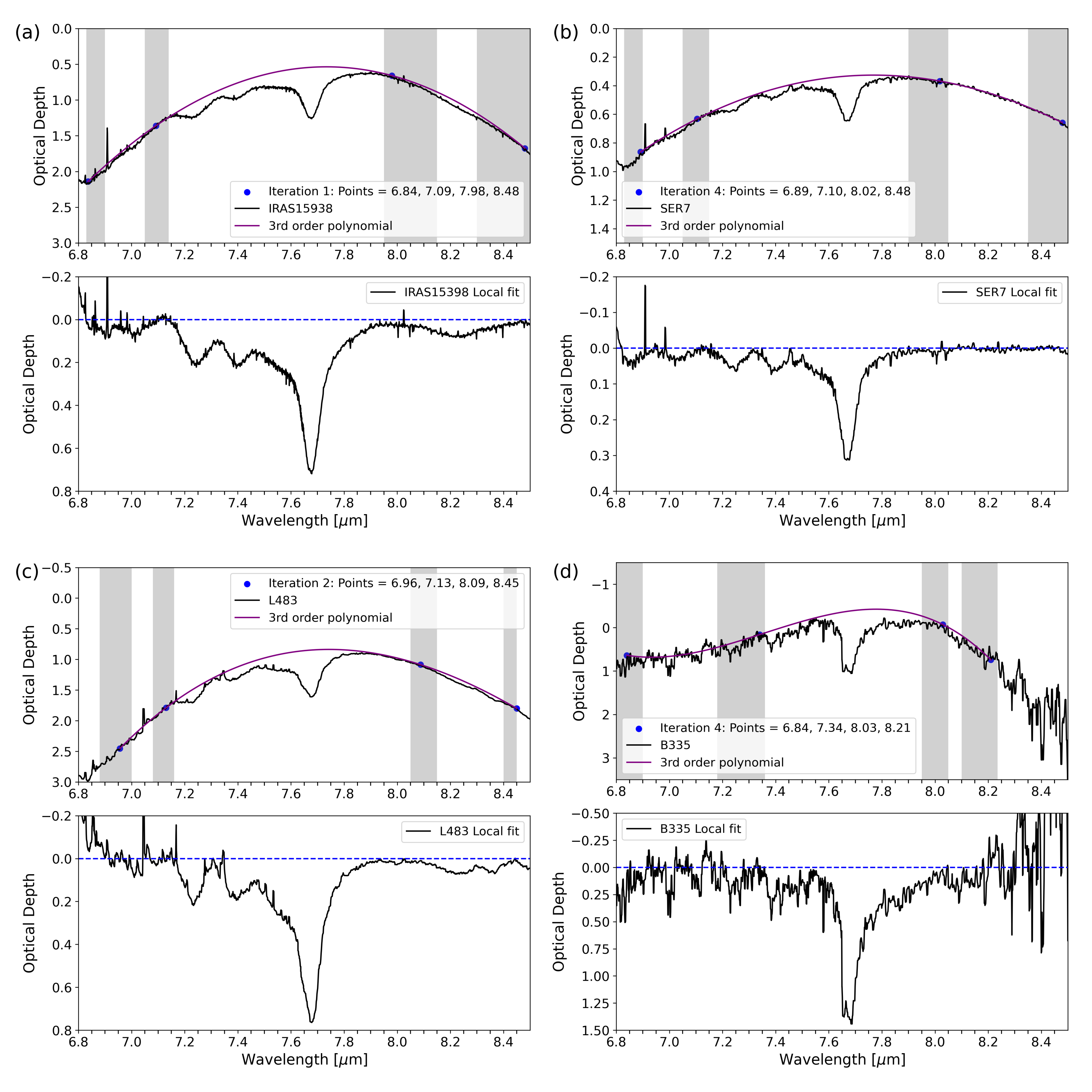}
  \caption{Example of baseline fits for each of the four CORINOS sources. The top plot with the vertical grey bars show the areas of minimal {\em local} ice absorption. This choice assumes that large absorption features (e.g., due to water or silicates) smoothly vary over our wavelength range that is fit by our polynomial. For each protostar one example set of guiding points is shown as blue markers, with the corresponding polynomal fit as a purple line. The resulting local optical depth spectra is shown below each panel. This process is repeated fifty times to generate different baseline subtracted optical depth spectra.} 
  
  \label{baseline_gallery}
\end{figure*}

Next, we fit a local baseline to isolate the absorption specific to the blended \sot\ and \methane\ feature.
It is common in the literature to identify ``by eye'' a single, fixed polynomial continuum, making it difficult to assess how sensitive the derived column density is to this decision. To quantify the impact of choice of baseline, we introduce a statistical technique to estimate the baseline-attributed uncertainty. Within the 6.8–8.5~$\mu$m spectral range near the \sot\ feature, we identify smooth wavelength regions that are free of strong {\em local} ice absorption. We use these regions to estimate the variation in possible polynomial fits. For each estimate of the baseline, a single point is randomly selected within each of the spectral regions and used as a guide point to fit a third order polynomial. From preliminary testing, a third-order polynomial was determined to best fit the local continuum for all four sources.  The wavelength regions identified for fitting \sot\ are marked as vertical grey bands in Figure \ref{baseline_gallery}, along with an example of the baseline fitting procedure and the resulting optical depth.

We followed the same procedure for each of our sources, except in the case of B335, which was substantially more noisy than the other sources. To keep the spectral guide points from falling into noise troughs when they were randomly selected, we took the median optical depth within each grey band for B335. We defined a $\pm10$ $\%$
window within that median value which allowed us to avoid noisy deviations. This also allowed us to avoid excessive smoothing that could alter the shape of the spectral features. We also note that in panel (d) of Figure~\ref{baseline_gallery}, that the spectrum of B335 is very noisy red-ward of 8.2~$\mu$m. To avoid aggressive smoothing of the spectra past this wavelength we chose to limit the red side of our spectral fitting zone to 8.2~$\mu$m.

\begin{figure*}
  \centering
  \includegraphics[scale=0.57]{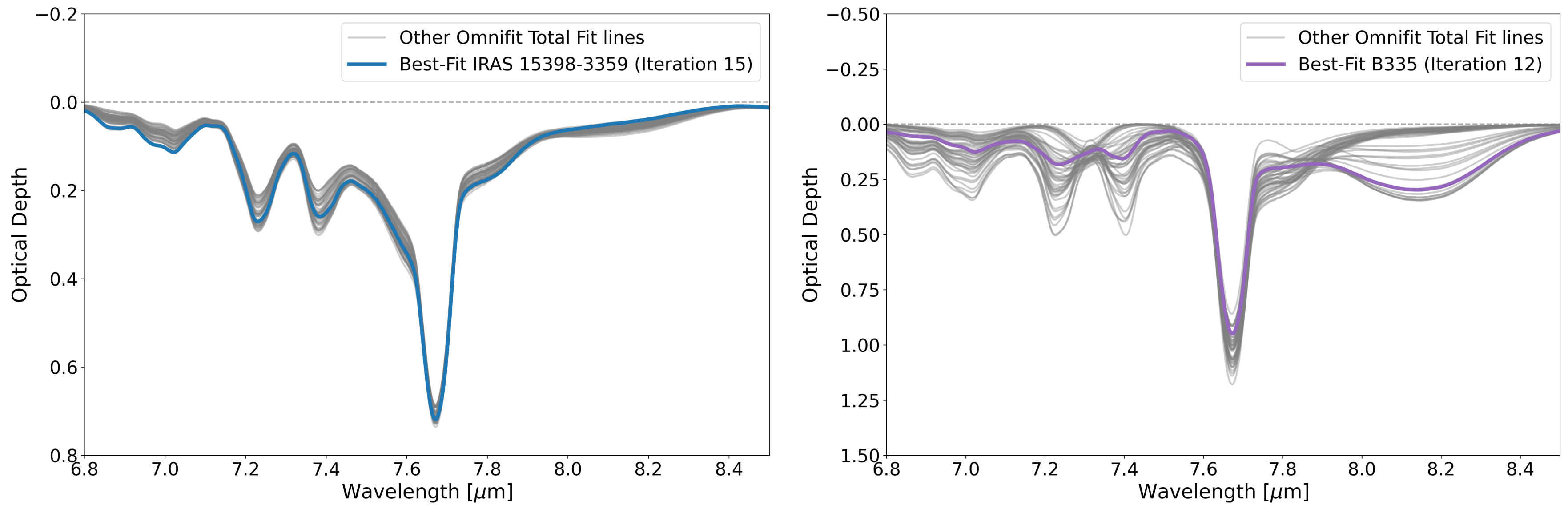}
  \caption{The colored line in each panel is the ''best-fit'' total \texttt{Omnifit} result out of the 50 baseline iterations using Reduced $\chi^2$ for the pure \sot\ ice. The grey lines represent all other total fit lines. On the left panel, we show IRAS 15398-3359 as an example of a source with little variation in the fitting routine compared to B335 in the right side panel, whose fitting varies substantially with baseline choice. } 
  
  \label{best_fit}
\end{figure*}

We repeat this randomized selection of polynomial anchor-points and fitting procedure fifty times for each source, producing a total of fifty optical depth spectra, referred to as ``iterations'' throughout this paper. The choice of fifty iterations reflects statistical completeness, providing sufficient sampling of baseline variability without oversampling redundant realizations. To visualize the effect of baseline variability, we selected two representative sources to show the outcome of the 50 randomized iterations (Figure~\ref{best_fit}): IRAS 15398–3359, which shows minimal variation across iterations, and B335, which exhibits substantially larger variation. These examples illustrate the extremes of baseline sensitivity in our dataset and provide a sense of the uncertainty range introduced by the fitting procedure.

This approach provides a statistical framework for assessing the uncertainties on the measured optical depth and thus column density introduced by baseline placement. Although some dependence on the adopted continuum windows and polynomial order remains, this method substantially reduces subjectivity compared to a single user-defined baseline.

\begin{table*}[ht!]
\centering
\caption{Laboratory Ice Mixtures and Band Strengths used in Spectral Fitting}
\label{lab_data}
\setlength{\tabcolsep}{2pt}       
\begin{tabular}{c c c c c c c l}
\hline\hline
Structure & Formula & Name & Ice Mixture & $\lambda$ ($\mu$m) & A (cm molec\textsuperscript{-1}) & Temp (K) & Ref. \\
\hline

\includegraphics[width=0.04\textwidth]{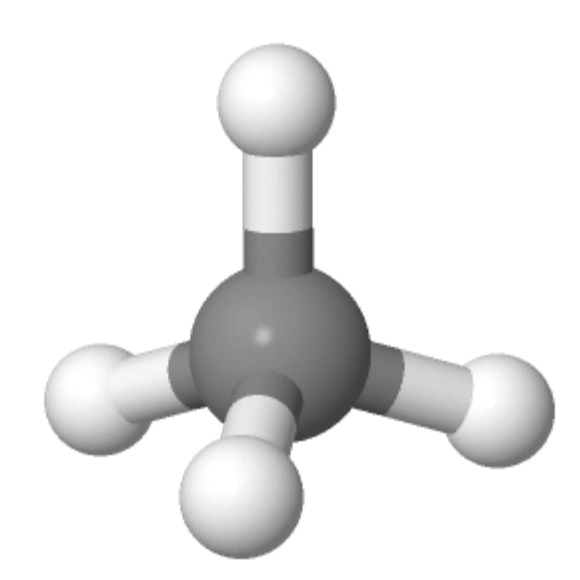}
 & CH$_4$
 & Methane
 & \parbox[t]{3.3cm}{H$_2$O:CH$_3$OH:CO$_2$:CH$_4$}
 & 7.67
 & \parbox[t]{2.5cm}{$9.6\times10^{-18}$}
 & 10
 & [1] \\

\includegraphics[width=0.04\textwidth]{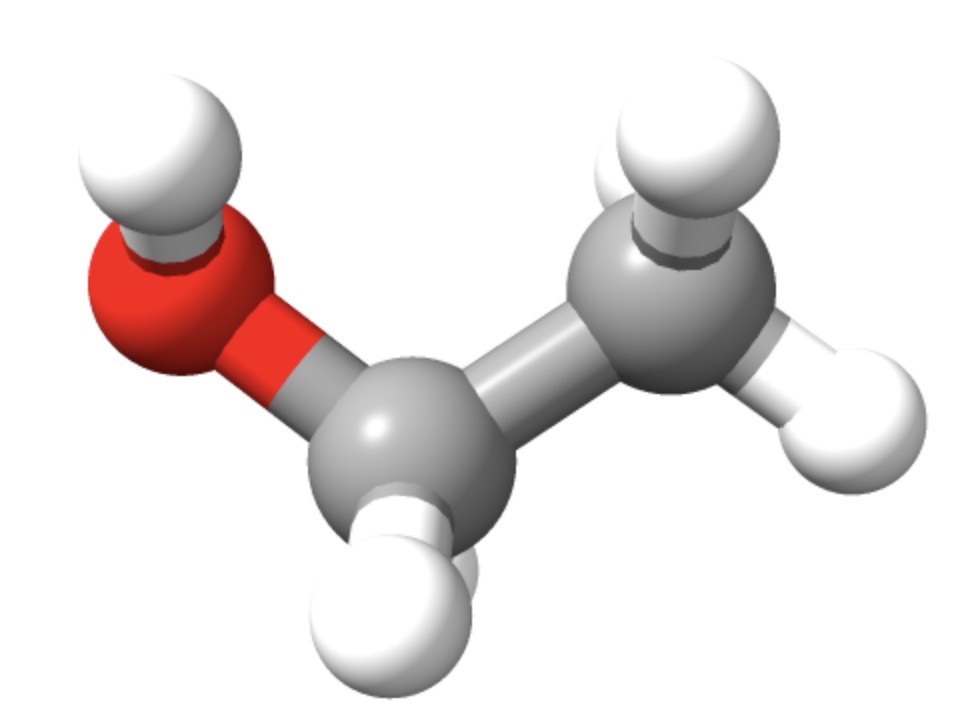}
 & CH$_3$CH$_2$OH
 & Ethanol
 & \parbox[t]{3.3cm}{H$_2$O:CH$_3$CH$_2$OH (20:1)}
 & 7.01, 7.47
 & \parbox[t]{2.5cm}{$3.0\times10^{-17}$}
 & 15
 & [2] \\

\includegraphics[width=0.04\textwidth]{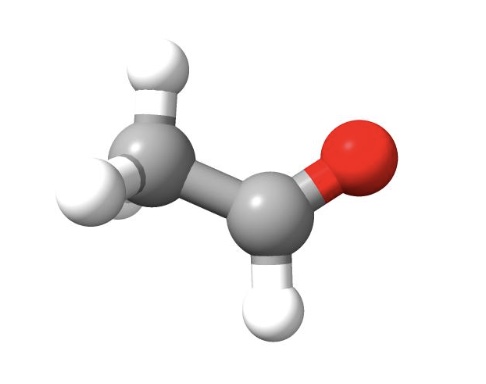}
 & CH$_3$CHO
 & Acetaldehyde
 & \parbox[t]{3.3cm}{H$_2$O:CH$_3$CHO (20:1)}
 & 7.23
 & \parbox[t]{2.5cm}{$1.4\times10^{-17}$}
 & 15
 & [2] \\

\includegraphics[width=0.04\textwidth]{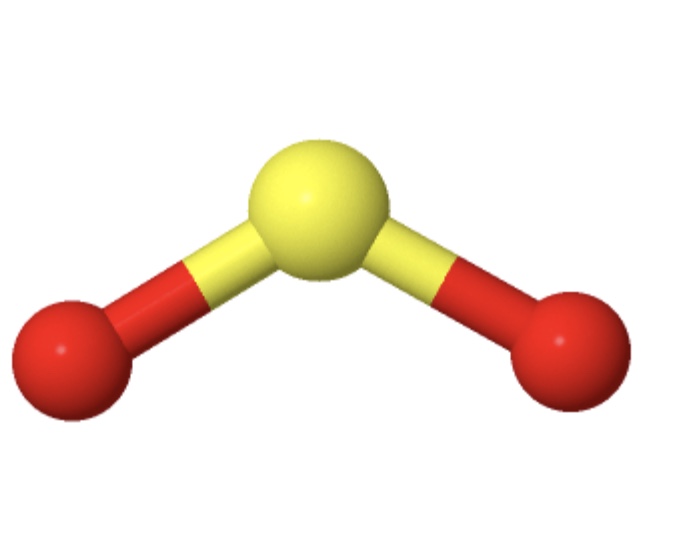}
 & SO$_2$
 & Sulfur dioxide
 & \parbox[t]{3.3cm}{pure, SO$_2$:CH$_3$OH (1:1)\\
                    and SO$_2$:H$_2$O (1:10)}
 & 7.60
 & \parbox[t]{2.5cm}{$3.4\times10^{-17}$,\\ $3.92\times10^{-17}$}
 & 10
 & [3,4] \\

\includegraphics[width=0.04\textwidth]{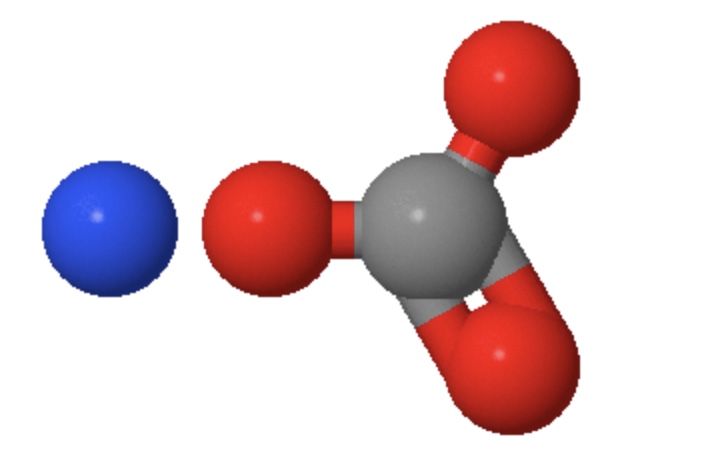}
 & NH$_4$COOH
 & Ammonium formate
 & \parbox[t]{3.3cm}{H$_2$O:NH$_4$COOH (7:100)}
 & 7.22, 7.41
 & \parbox[t]{2.5cm}{$1.0\times10^{-17}$}
 & 14
 & [5] \\

\includegraphics[width=0.04\textwidth]{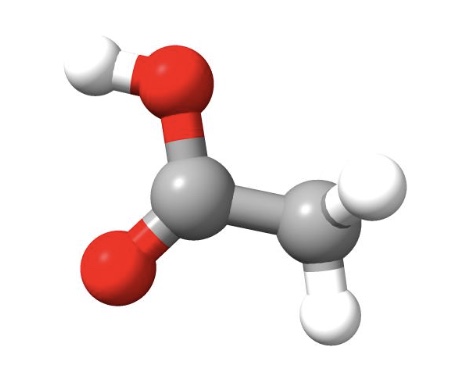}
 & CH$_3$COOH
 & Acetic acid
 & \parbox[t]{3.3cm}{H$_2$O:CH$_3$COOH (10:1)}
 & 6.83, 6.89
 & \parbox[t]{2.5cm}{$1.9\times10^{-17}$}
 & 10
 & [6] \\

\includegraphics[width=0.04\textwidth]{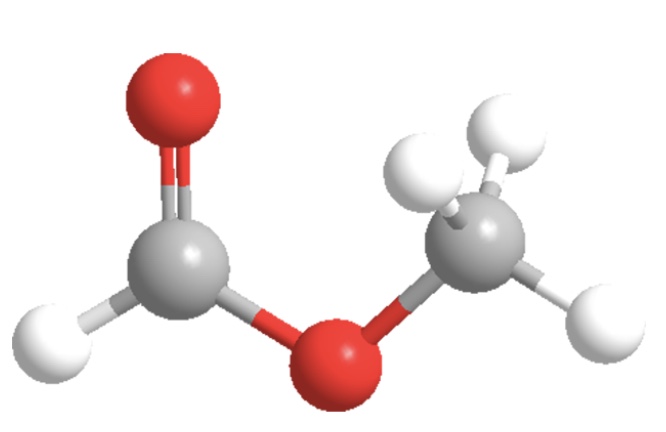}
 & HCOOCH$_3$
 & Methyl formate
 & \parbox[t]{3.3cm}{H$_2$O:HCOOCH$_3$ (20:1)}
 & 7.75
 & \parbox[t]{2.5cm}{$2.52\times10^{-17}$}
 & 15
 & [7] \\

\includegraphics[width=0.04\textwidth]{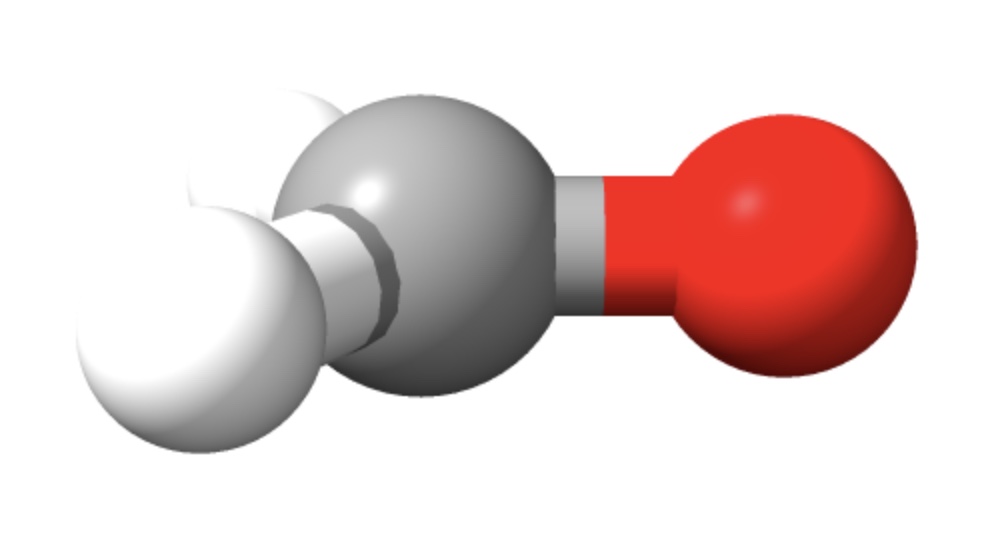}
 & H$_2$CO
 & Formaldehyde
 & \parbox[t]{3.3cm}{pure H$_2$CO}
 & 7.75
 & \parbox[t]{2.5cm}{$1.4\times10^{-17}$}
 & 10
 & [8] \\

\includegraphics[width=0.04\textwidth]{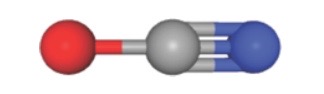}
 & OCN$^-$
 & Cyanate ion
 & \parbox[t]{3.3cm}{pure OCN$^-$}
 & 7.6, 4.6
 & \parbox[t]{2.5cm}{$7.45\times10^{-18}$,\\ $1.3\times10^{-16}$}
 & 12
 & [9,10] \\

\includegraphics[width=0.04\textwidth]{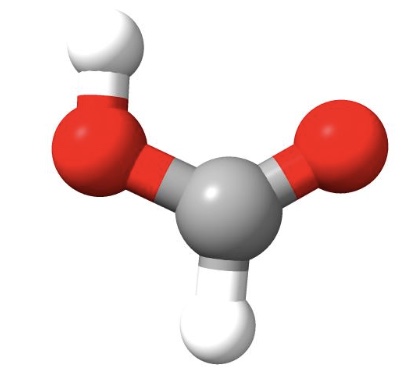}
 & HCOOH
 & Formic acid
 & \parbox[t]{3.3cm}{HCOOH:H$_2$O:CO (8:62:30),\\ HCOOH:CO (11:89)}
 & 8.25, 4.67
 & \parbox[t]{2.5cm}{$2.9\times10^{-17}$}
 & 15
 & [2,6] \\

\includegraphics[width=0.03\textwidth]{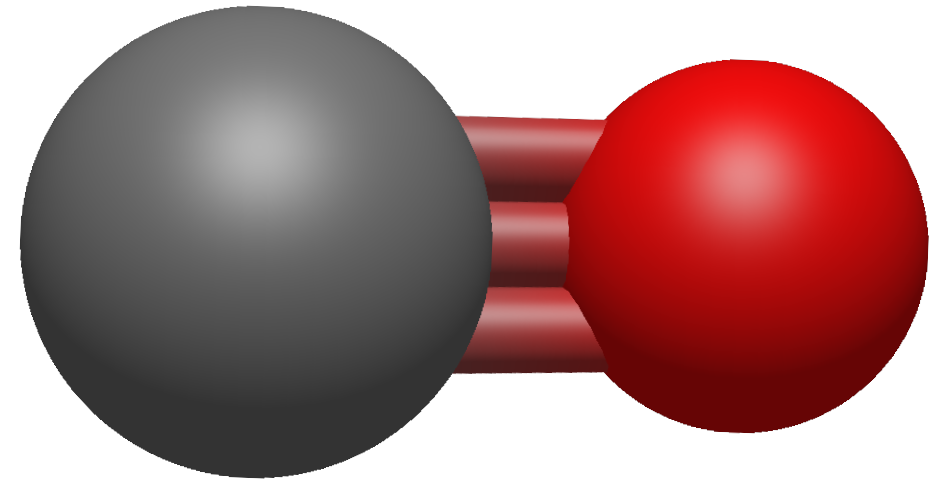}
 & CO
 & Carbon monoxide
 & \parbox[t]{3.3cm}{CO:CH$_3$OH: CH$_3$CH$_2$OH\\ (20:20:1)}
 & 4.67
 & \parbox[t]{2.5cm}{$1.4\times10^{-17}$}
 & 15
 & [2] \\
\hline
\end{tabular}
\vspace{2mm}

\textbf{References.}
[1] \citet{Kerkhof1999};
[2] \citet{TvS2018};
[3] \citet{Boogert1997};
[4] \citet{Yarnall2022};
[5] \citet{Galvez2010};
[6] \citet{Bisschop2007};
[7] \citet{TvS2021};
[8] \citet{gerakines1996};
[9] \citet{rocha2024};
[10] \citet{vanBroekhuizen2005}.\\
{\textbf{Note:} Data sourced from Leiden Ice Database for Astrochemistry \citep[LIDA,][]{rocha2022LIDA}, and the Cosmic Ice Laboratory database for reference [4]. The table shows the main species targeted in the 6.8 to 8.5~$\mu$m window, and whether each species was pure or mixed, along with ratios. Species that were used in fitting the OCN$^-$ feature at 4.6~$\mu$m are also listed.}
\end{table*}

\subsection{Laboratory Data and \texttt{Omnifit}}\label{labdata} 
To determine ice column densities from the MIRI spectra, we fit each observed optical depth spectrum with laboratory ice spectra from the Leiden Ice Database for Astrochemistry (LIDA) \citep{rocha2022LIDA}. The chosen lab spectra (summarized in Table \ref{lab_data}) cover the relevant 6.8-8.5~$\mu$m absorption features. The laboratory spectra include both pure and mixed ices designed to reproduce the molecular environments expected in interstellar and protostellar grain mantles. Mixed ices are prepared by combining two or more species under controlled conditions to examine how the surrounding matrix influences the shape and strength of absorption features. Each entry lists the constituents of the ice mixture and their corresponding ratios, the adopted band strength ($A$, cm molecule$^{-1}$) used to convert integrated optical depth to column density, and the temperature in Kelvin (K) at which the spectrum was measured in the laboratory.

The raw laboratory spectra are provided in absorbance units (Abs$_\nu = -\log{10}(I/I_0)$), so we first convert them to optical depth using $\tau_\nu = ln(10){Abs}_\nu$) in order to directly compare with the observed $\tau_\nu$ for each source. Each laboratory spectra is then interpolated onto the wavelength grid of the observations and smoothed to the JWST/MIRI spectral resolution. Because the lab data were measured with Fourier-transform infrared (FTIR) spectroscopy, some spurious interference fringes or baseline curvature can be present. We correct these by fitting a low-order polynomial, typically 2nd or 4th order, to the continuum baseline of the lab spectrum and subtracting it, producing a flattened baseline. This is an important step as any spurious features associated with the lab data could be misinterpreted when fitting to the observed spectrum. This results in a processed lab optical depth spectrum for each ice species that can be matched to the observational data. 

Although our primary goal is to characterize the SO$_2$ and CH$_4$ ice features, we still must account for overlapping absorption from different ice components that have local features in the same wavelength range. In particular, we consider CH$_3$CH$_2$OH (ethanol), HCOOH (formic acid), NH$_3$:HCOOH mixtures which produce the anion HCOO$^-$ upon acid–base reaction \citep[e.g.,][]{Novozamsky2001}, OCN$^{-}$ (cyanate ion), CH$_3$CHO (acetaldehyde), H$_2$CO (formaldehyde), HCOOCH$_3$ (methyl formate), and CH$_3$COOH (acetic acid) in addition to the target SO$_2$ and CH$_4$ (see Table \ref{lab_data} for a complete list). This choice of additional ice components is based off preliminary trial and error with fitting laboratory data as well as guided by previous studies to fit this region showing that multiple components are likely to contribute to fitting the fingerprint region \citep{Boogert1997,yang2022,mcclure2023ices, rocha2024, Rocha2025}. Including these species in the model ensures that absorption near the SO$_2$ and CH$_4$ bands are properly fit by their likely carriers. \citet{rocha2024} found that ten molecular ice components (including CH$_4$, SO$_2$, and the organics and ions listed above) are required to reproduce the MIRI spectra of protostars in this range, with CH$_4$ dominating the 7.7~$\mu$m band and species like OCN$^-$, HCOO$^-$, C$_2$H$_5$OH, and CH$_3$CHO contributing to the 7.2 and 7.4~$\mu$m features. Previous studies had also attributed the 7.24 and 7.41~$\mu$m ISO/Spitzer bands to combinations of HCOO$^-$, C$_2$H$_5$OH, and CH$_3$CHO \citep{Schutte1999, oberg2011sp}. Guided by this prior work, our fitting includes the most relevant ice candidates for the 6.8-8.5~$\mu$m region. We also examine how the fitted CH$_4$ column density varies under different baseline fits to provide context for the relative sensitivity of the SO$_2$ feature.

The spectral decomposition of the CORINOS sources using the laboratory data in Table \ref{lab_data} was carried out using \texttt{Omnifit}\footnote{\href{https://doi.org/10.5281/zenodo.33056}{Omnifit} is an open‑source Python library designed to interpret observational IR spectra.} and performs non‑linear least‑squares optimization via the \texttt{lmfit}\footnote{\url{https://lmfit.github.io/lmfit-py/}} package to scale and fit multiple lab spectra to best match the observed absorption profile. The optimization routine developed by \cite{Suutarinen2015_thesis} is the Levenberg–Marquardt algorithm via \texttt{lmfit} that minimizes the following simplified expression:

\begin{equation}
\chi^2(\vec{a}) = \sum_{i=1}^N \left( \frac{y_i - f(x_i; \vec{a})}{\sigma_i} \right)^2
\end{equation}
\noindent $y_i$ is the observed optical depth at wavelength $i$ out of $N$ data points, and $f(x_i; \vec{a})$ is the model's predicted optical depth evaluated at the same wavelength $x_i$. The parameter $\vec{a}$ are the scale factors that weight each laboratory spectrum. \texttt{Omnifit} starts with an initial input guess of scaling factor (0.0) and then finds the set of scalings until it minimizes $\chi^2$($\vec{a}$).

\begin{figure*}[htbp]
  \centering
  \includegraphics[width=0.46\textwidth]{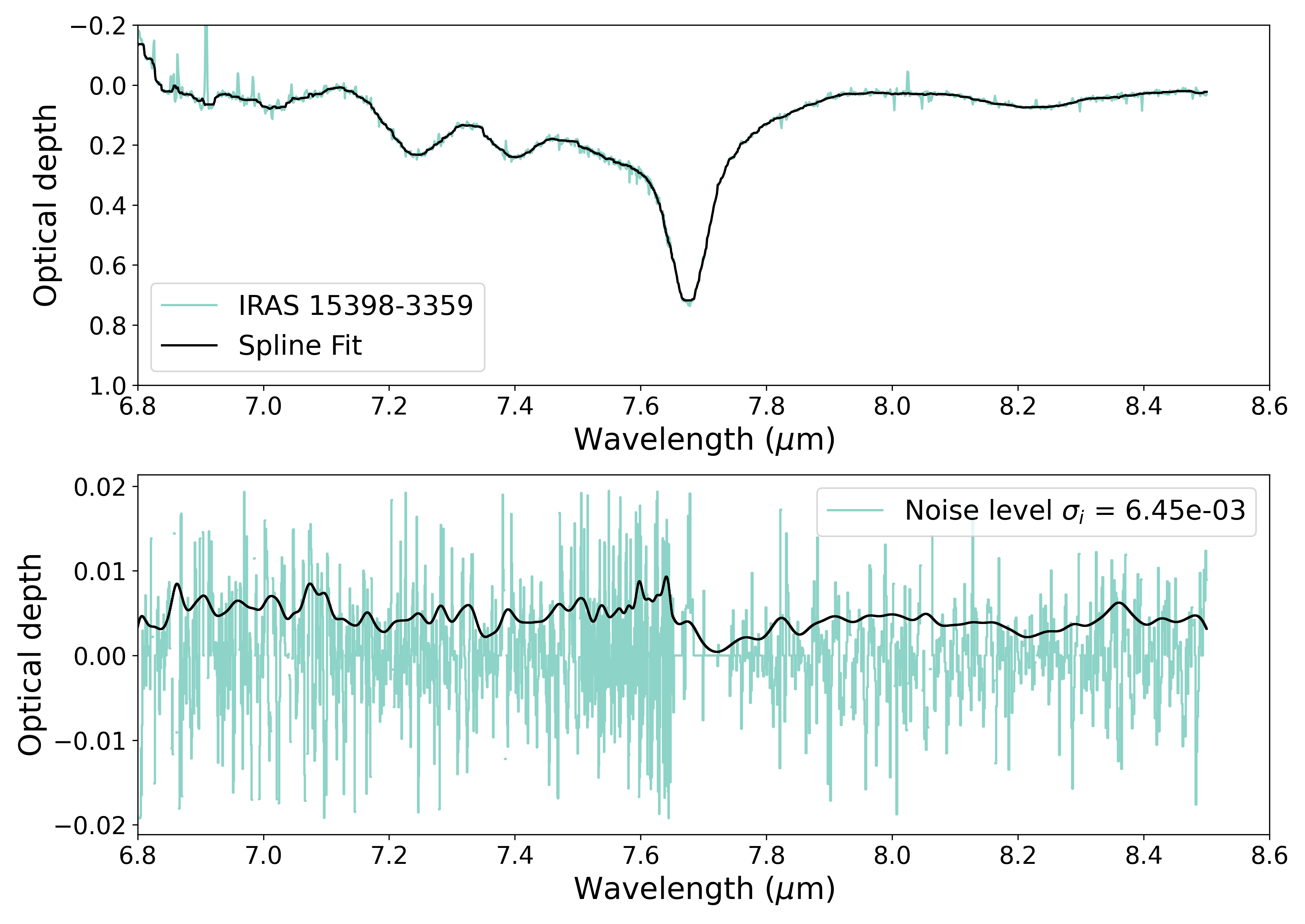}%
  \hspace{0.05\textwidth}%
  \includegraphics[width=0.46\textwidth]{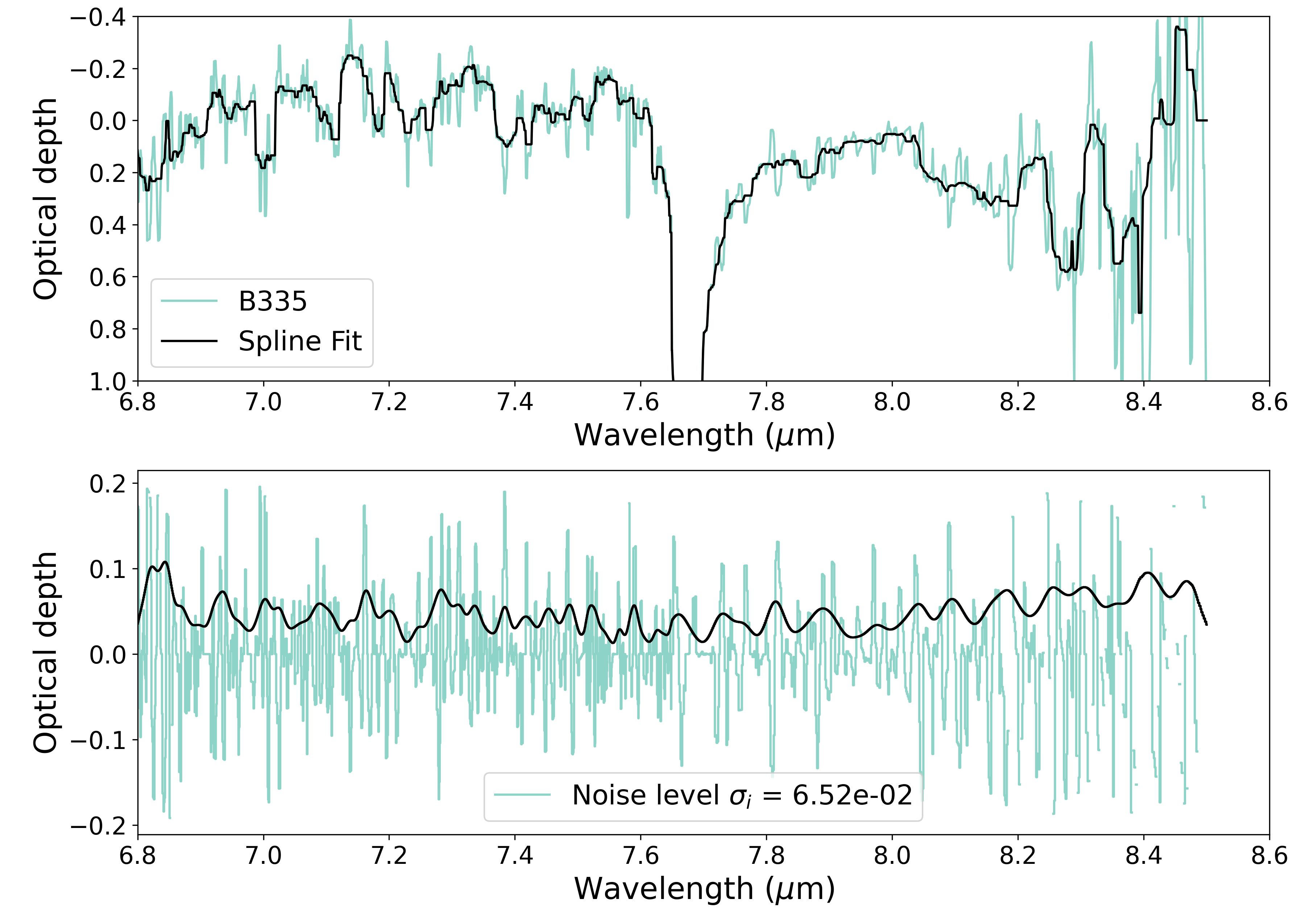}
  \caption{Example of fitting a spline to the local optical depth spectrum of IRAS 15398-3359, with the residuals used to derive the per-point noise level $\sigma_i$ for \texttt{Omnifit}. Right: The same procedure is applied to B335, which yields a noticeably higher noise level.}

  \label{sidebyside_noise}
\end{figure*}

The value $\sigma_i$ is the per-point measurement uncertainty that is passed into the fitting routine, representing the random scatter remaining in each baseline-subtracted spectrum and serving as our estimate of the effective noise floor. To determine this quantity, we apply a 21-point kernel median filter to the optical-depth spectrum to produce a smooth continuum, then subtract it to obtain the residuals. A 3-$\sigma$ clipping is used to reject outliers, and the standard deviation of the remaining residuals defines $\sigma_i$. In \texttt{Omnifit}, this value is supplied as the constant error array so that each wavelength channel is weighted equally according to the measured noise level, $\sigma_i$. Figure \ref{sidebyside_noise} illustrates this procedure for IRAS 15398-3359 and B335, showing that the latter exhibits a noticeably higher noise level.

This standard deviation is used as a single, constant $\sigma_i$ that \texttt{Omnifit} uses as the uncertainty at each wavelength ($y_i$). This allows the Levenberg–Marquardt algorithm to properly scale the contributions of each wavelength according to the actual noise level in the data. It will return both the best‑fit amplitudes and their 1~$\sigma_i$ errors from the covariance matrix.
After fitting all the laboratory components \texttt{lmfit} reports the total $\chi^2$, as well as the reduced $\chi^2$ which gives an average squared normalized residual per independent degree of freedom. The reduced $\chi^2$ is used to determine the ``best-fit" out of all the iterations for each source.

We verified that the features of SO$_2$ and CH$_4$ were being well reproduced once the nearby absorption features from species like formate and acetaldehyde were taken into account. In cases where a species was clearly not contributing, it was dropped from the final fit to reduce parameter degeneracy. The final \texttt{Omnifit} solution yields a scaled optical depth profile for each ice component, from which we can determine column densities. For each ice species, the column density $N_{\rm ice}$ molecules cm$^{-2}$ is derived from the fitted optical depth spectrum using the equation:

\begin{equation}
N_{\mathrm{ice}} = \frac{1}{A} \int_{\nu_1}^{\nu_2} \tau_\nu^{\mathrm{lab}} \, d\nu ,
\label{colden}
\end{equation}

\noindent where $A$ is the band strength of the corresponding vibrational mode (in units of cm molecule$^{-1}$). The integral represents the integrated optical depth area of the absorption band from $\nu_1$ to $\nu_2$, covering the frequency range on either side of the specific wavelength for each species in Table \ref{lab_data}. $A$ quantifies how much IR light a molecule absorbs at that wavelength and is measured in the laboratory by calibrating the absorbance with a known column density. The band strength $A$ is not a universal constant for a given molecule, but can depend on the ice conditions such as temperature, phase, and mixture with other species \citep[e.g.,][]{Oberg2007icemixtures}. Choosing lab spectra close to astrophysical ice conditions is important for obtaining accurate column densities. 

The intensity of the band strength for the deformation mode of \methane\ at 7.7 ~$\mu$m increases when it is highly diluted in a mixture, but in a more 1:1 ratio the band strength is roughly 21-23 $\%$ stronger. 
\water\ is the most abundant ice in protostars, and is often the first to freeze onto dust grains due to its high sublimation temperature. \water\ therefore provides a matrix from which other species form, such as \methane\ \citep{Qasim2018}. Multiple compositions for testing mixtures of these species was used to best fit the 7.7 ~$\mu$m feature, and the best fit for our study was H$_2$O:CH$_3$OH:CO$_2$:CH$_4$ (0.4:0.6:1.0:0.23). The band strength for this mixture is $9.6\times10^{-18}$ cm molecule$^{-1}$ from \cite{Kerkhof1999}. The \sot\ band strength is less sensitive to its environment, and so the same value can be used for the pure and \methanol:\sot\ ice. For the \water\ mixed \sot\ ice, we used a different band strength as water is the dominant component in the 10:1 laboratory ice and has a greater affect \cite{Yarnall2022}. We have used the same care in selecting the band strength values for the other species in our fits and the $A$ values adopted in this work are listed in Table \ref{lab_data}.

The mixture composition of the laboratory spectra not only affects the band strengths, but also the shapes and positions of the IR peaks. Species isolated in a pure ice can have different peak positions and intensities than when they are diluted in H$_2$O or CO$_2$ matrices \citep{Ehrenfreund1997}. We therefore used relevant laboratory mixtures when those are thought to better represent the interstellar ice matrix. Decades of laboratory work by various groups have produced an extensive archive of such spectra, including studies by \citet{Boogert1997, Gerakines1995, Oberg2007icemixtures, Ehrenfreund1996}, among others. Many of these are now accessible through databases such as LIDA \citep{rocha2022LIDA}. Utilizing these resources and the guidance from recent JWST ice studies \citep[e.g.,][]{mcclure2023ices, rocha2024}, we were careful to include lab spectra that captures the most realistic combinations of mixtures for the ices in our sample.

The choice of laboratory ice spectra should also reflect the physical state of the observed ices. At low temperatures, ices are amorphous and display broad absorption features, while heating can induce crystallization, producing sharper and more structured IR bands \citep{Dartois2015, Gudipati2023}. The most widely used diagnostic of this process is the 3~$\mu$m H$_2$O stretching band in the near infrared or the 6~$\mu$m bending mode in the mid infrared, which shows clear structural changes with increasing temperature \citep{Boogert2015}. Additional heating diagnostics include the CO$_2$ bands: the 4.3~$\mu$m $^{13}$CO$_2$ stretching and 15.2~$\mu$m bending modes, both of which undergo characteristic profile changes as ices crystallize. Such features are especially clear in IRAS 15398–3359, as also noted by \citet{yang2022}. Indeed, Kim et al. (in prep) find that fitting the global JWST MIRI spectrum of IRAS 15398–3359 requires at least two distinct H$_2$O ice temperature components to reproduce the observed profiles, suggesting a complex thermal history. This type of fitting also provides insight into the temperature evolution of the observed ices. By contrast, volatile organics such as CH$_3$OH, CH$_3$CHO, and SO$_2$ show only minor peak shifts or bandwidth changes at low temperatures (10–15 K). Significant profile changes emerge only once ices are warmed above $\sim$70~K \citep{Oberg2009, Rachid2020, TvS2021}. For the 6.8–8.5~$\mu$m region analyzed here, we find that low-temperature (10–15~K) laboratory spectra provide the best match to the observed features in our CORINOS sample.

\section{\texttt{Omnifit} Results} \label{results}
\subsection{CH$_4$ Ice Column Density} 
The prominent 7.7~$\mu$m absorption feature is clearly detected in all four protostellar spectra and is attributed to the $\nu_4$ bending mode of CH$_4$ ice. This methane band is the main carrier in the 6.8-8.5~$\mu$m region \citep{oberg2008}.
IRAS 15398-3359 shows the CH$_4$ feature at an integrated $\tau_{\rm CH_{4}} \sim 7~$cm$^{-1}$ (peak $\tau \sim 0.5$), consistent with previous studies of low-mass protostars \citep{rocha2024}. These values are higher than the previous fits of \citep{oberg2008}, which found $\tau_{\rm CH_{4}} \sim 4.2\pm0.8~$cm$^{-1}$ using Spitzer IRS observations for this same source. The primary difference between the previous study and this one is that at the lower resolution (R$<$100), the baseline could not account for ``wiggles'' that we now resolve as ice features with JWST in this wavelength range(see their Figure 1), and so the previous baseline had been too deep near CH$_4$, reducing the calculated ice optical depth. L483 shows a similar integrated band area (integrated $\tau_{\rm CH_{4}} \sim7~$cm$^{-1}$, peak $\tau_{\rm CH_{4}} \sim0.5$), while Ser-emb 7 is slightly weaker by roughly 20\%. In contrast, B335 exhibits a much stronger CH$_4$ absorption (integrated $\tau_{\rm CH_{4}}$ of 12 cm$^{-1}$ and a peak $\tau_{\rm CH_{4}} \sim1$). From these fits, we derive CH$_4$ column densities on the order of $10^{17}-10^{18}$~cm$^{-2}$ for the four sources (see Table \ref{ratios}). We note that the B335 spectrum approaches the noise floor across much of the 7–8 $\mu$m region. While we report derived column density values for completeness, they should be regarded as tentative measurements.

Across the different baseline fits, the derived CH$_4$ column densities remain relatively consistent, indicating that the feature is robust against moderate continuum variations. This stability provides an empirical reference for evaluating the greater baseline sensitivity of the weaker SO$_2$ absorption.

\subsection{OCN$^-$ Column Densities from NIRSpec} \label{ocn_results}

In order to more definitively determine the contribution of OCN$^-$ to the 7.6~$\mu$m feature, it is necessary to use the 4.6~$\mu$m band in the near infrared. We use a simplified baseline fitting procedure here compared to our Monte Carlo approach applied to the MIRI data in this paper. The NIRSpec data for both B335 and IRAS~15398$-$3359 were used to estimate the column density for the 4.6~$\mu$m OCN$^-$ band. The B335 data are from the JWST IPA program (ID: 1802; \citealt{Federman2024}), while the IRAS~15398-3359 data are from JWST program 1854 (PI: McClure). To our knowledge, no published NIRSpec observations of the 4.6~$\mu$m OCN$^-$ band exist yet for L483 or Ser-emb~7. Without this near-IR constraint, the OCN$^-$ column density cannot be reliably estimated from the 7.6~$\mu$m MIRI feature for those sources. As a result, a direct NIRSpec–MIRI comparison is only possible for B335 and IRAS~15398-3359, and we therefore restrict our OCN$^-$ analysis in this section to these two objects.

\begin{deluxetable*}{lcccc}
\tablecaption{OCN$^-$ Column Densities and Adopted Band Strengths\label{ocn_compare}}
\tablehead{
  \colhead{Source} &
  \colhead{$\lambda_\mathrm{central}$\,($\mu$m)} &
  \colhead{$N(\mathrm{OCN}^-)$\,(cm$^{-2}$)} &
  \colhead{$A$\,(cm molecule$^{-1}$)} &
  \colhead{Reference}
}
\startdata
IRAS\,15398$-$3359 & 4.62 & $1.20\times10^{17}$ & $1.30\times10^{-16}$ & \citet{vanBroekhuizen2005} \\  
                   & 7.62 & $3.58\times10^{17}$ & $7.45\times10^{-18}$ & \citet{rocha2024}      \\[6pt]
B335               & 4.62 & $7.66\times10^{16}$ & $1.30\times10^{-16}$ & \citet{vanBroekhuizen2005}\\  
                   & 7.62 & $6.30\times10^{15}$ & $7.45\times10^{-18}$ & \citet{rocha2024}      \\
\enddata
\tablecomments{The 7.62~$\mu$m column densities were measured using the H$_2$O:\sot\ (10:1) laboratory mixture of \citet{Yarnall2022}.}
\end{deluxetable*}

\begin{figure}
  \centering
  \includegraphics[scale=0.34]{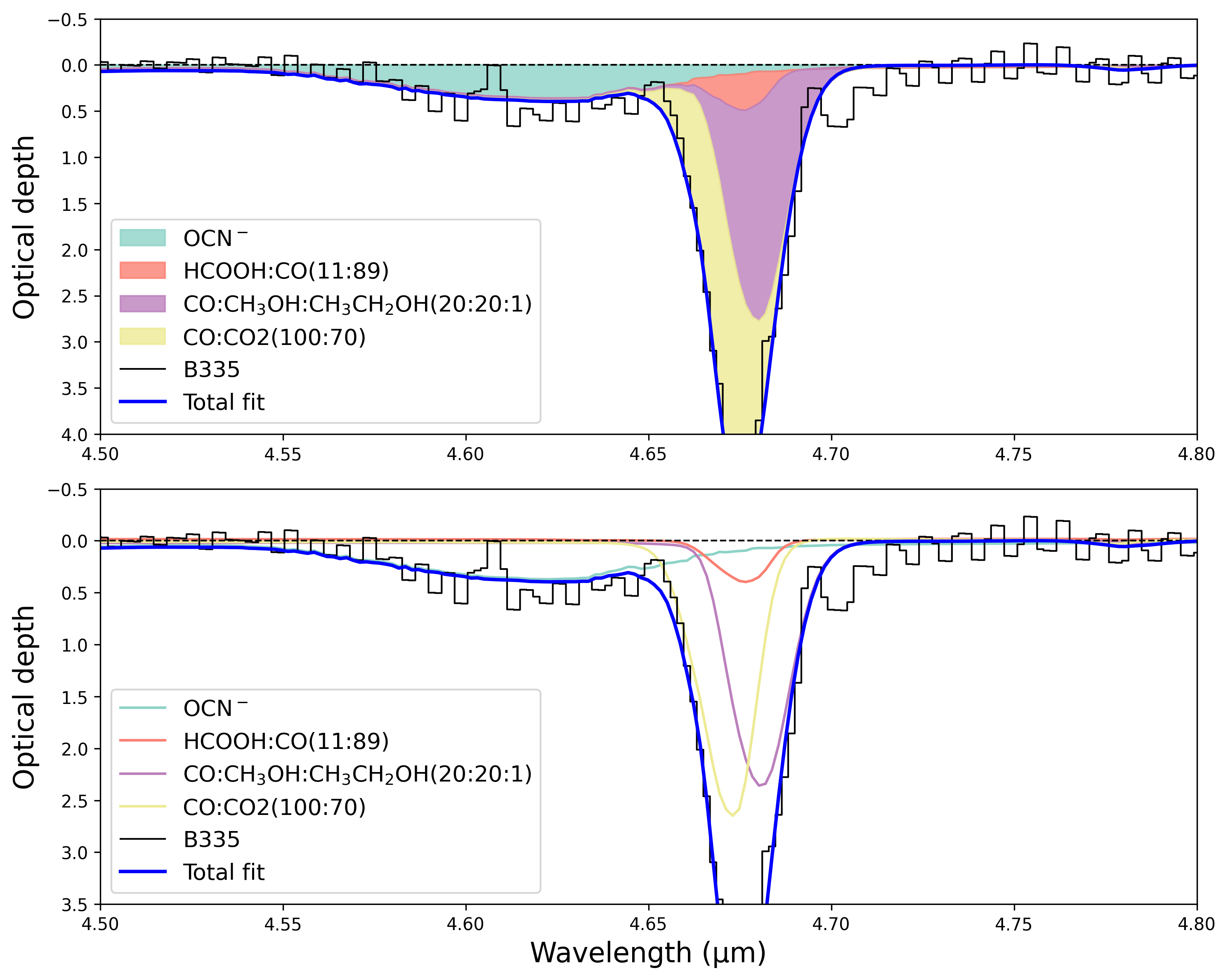}
  
  \caption{Fitting laboratory data to the 4.6 $\mu$m OCN$^-$ feature (teal) in NIRSpec data for source B335.}
  
  \label{B335_nirspec}
\end{figure}

To fit the baseline continuum for B335 we use a fourth order polynomial over ranges (4.0,4.2), (4.35,4.36), (4.51,4.52), (4.75,4.76), (4.8,4.85) to isolate the OCN$^-$ feature. We smooth the spectrum gently with a median filter kernel=3. The resulting \texttt{Omnifit} fitting results are shown in Figure \ref {B335_nirspec}, and the column densities in Table \ref {ocn_compare}. The 4.6~$\mu$m feature is blended with the CO stretching mode and therefore to properly fit this region, different mixtures of CO were tested.

We measure an OCN$^-$ column density of $7.66\times10^{16}$ cm$^{-2}$ in the 4.6~$\mu$m feature. Our result is higher than the $5.0\times10^{16}$ cm$^{-2}$ reported for this same source in \citet{Nazari2024_cyanides}, but the two values are broadly consistent when the reported uncertainties and differences in continuum subtraction, spectral smoothing and fitting methods are taken into account.

Interestingly, our MIRI data does not reproduce the same strength OCN$^-$ feature at 7.6~$\mu$m for the \water:\sot\ mixture, which had the best fit to the observed spectrum with only $6.30\times10^{15}$ cm$^{-2}$, one order of magnitude lower than the NIRSpec feature. Given multiple OCN$^-$ detections in the NIRSpec 4.6~$\mu$m feature, we would expect a correspondingly stronger 7.6~$\mu$m band in the MIRI data; however, the poor quality of the B335 MIRI spectrum prevents a fair comparison in this source.

\begin{figure}
  \centering
  \includegraphics[scale=0.34]{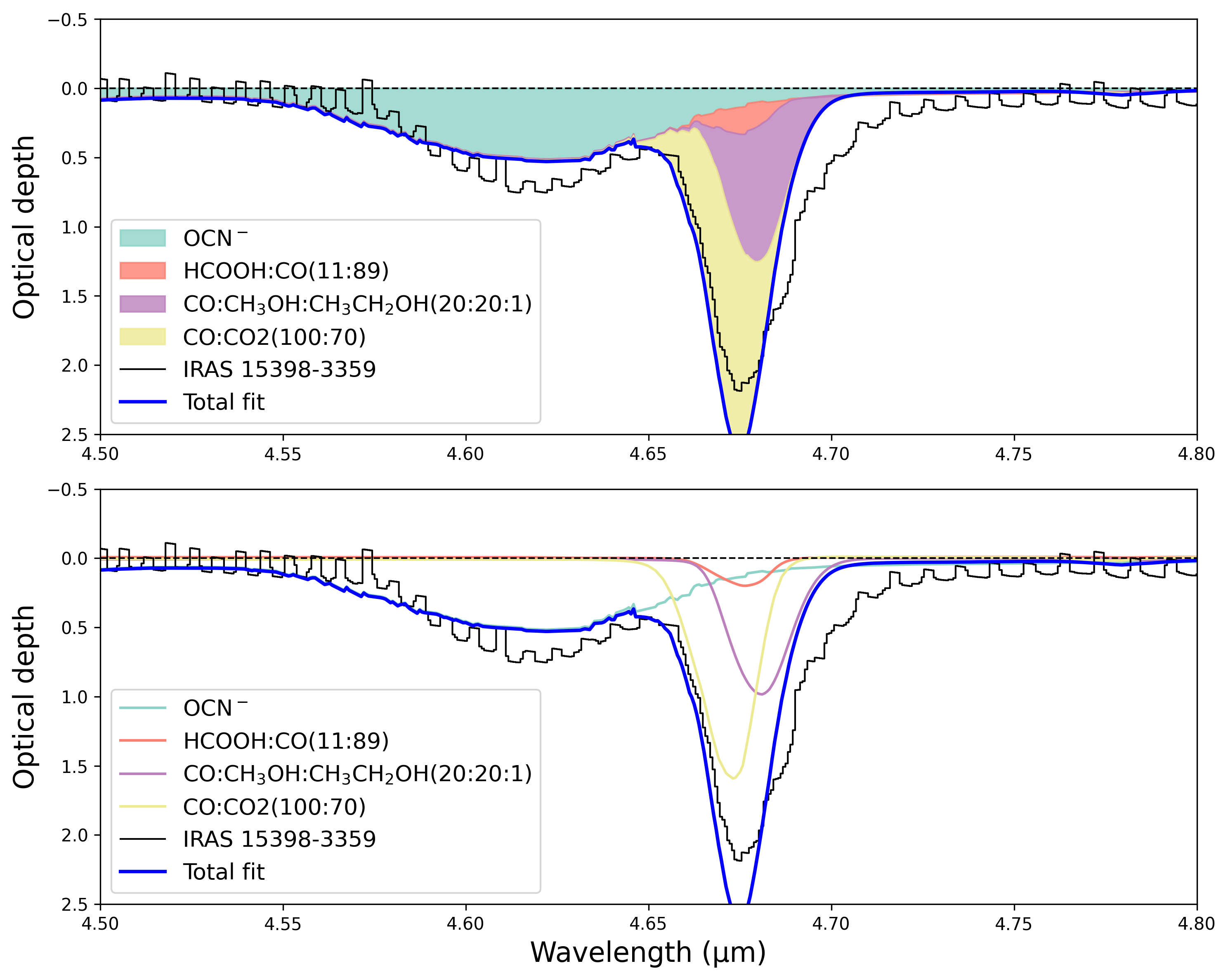}
  
  \caption{Fitting laboratory data to the 4.6 $\mu$m OCN$^-$ feature (teal) in NIRSpec data for source IRAS 15398-3359.}
  
  \label{iras_nirspec}
\end{figure}

\noindent

We followed the same procedure for IRAS 15398-3359's NIRSpec data. We used a third order polynomial over ranges (4.35,4.36), (4.51,4.52), (4.75,4.76), (4.8,4.85) to draw our baseline continuum, and then subsequently fit with \texttt{Omnifit}. The spectrum is smoothed with a median filter kernel=7 to decrease the CO gas emission lines prominent in this region. The results are shown in Figure \ref{iras_nirspec}. The CO stretching feature is broader on the redward side at 4.67~$\mu$m, and therefore does not fit as well as this same feature for B335, indicating that we have not found quite the right CO mixture to best fit this source. However, the OCN$^-$ feature is fairly well reproduced, and comparing to the column density measurement from the \water:\sot\ mixture in the MIRI data showed values for both the 4.6 and 7.6~$\mu$m features similar, with $1.20\times10^{17}$ cm$^{-2}$ and $3.58\times10^{17}$ cm$^{-2}$, respectively. 
A more detailed discussion of the implications of these OCN$^-$ measurements, including their role in the blended 7.6~$\mu$m region and comparison with previous studies, is presented in Section~\ref{ocn_discussion}. Importantly, our independent JWST NIRSpec detection of the 4.6~$\mu$m OCN$^-$ feature in IRAS 15398–3359 yielded a column density comparable to our derived value at the 7.6~$\mu$m feature when using the H$_2$O:SO$_2$ ice mixture, strengthening confidence in its presence in addition to \sot.

\subsection{SO$_2$ Column Density: Pure vs. Mixed Ices}\label{so2_fits}

On the blue-side of the CH$_4$ ice absorption there is a weak feature near 7.6 $\mu$m, which is attributed to the 
$\nu_3$ asymmetric stretching vibrational mode of \sot\ ice. Identifying \sot\ in absorption has been challenging as it is weak and blended with nearby features, including those from complex organics like ethanol at 7.24~$\mu$m and formate near 7.25~$\mu$m \citep{Boogert1997,Schutte1999}. 
\citet{Boogert1997} found that the spectral shape, or profile, of the 7.6 $\mu$m feature of \sot\ is highly sensitive to its environment, and noted that laboratory data of pure \sot\ or \sot\ mixed with \methanol\ ices provided a good fit for high-mass protostellar sources W33A and NGC 7538. \citet{rocha2024} found that the \sot:\methanol\ mixture provided the best fit in their study as well for NGC 1333 IRAS2A and IRAS 23385+6053. While species are rarely in a pure matrix in interstellar ices, for completeness we fit the data using one pure \sot\ ice laboratory spectra and two spectra of \sot\ mixed with CH$_3$OH ice (a \sot:CH$_3$OH mixture at 1:1 ratio) and mixed with H$_2$O \citep[a \water:\sot\ mixture at 10:1 ratio;][]{Yarnall2022}. Given the complex ice spectra we see in these data, the \sot\ ice mixtures are considered to be more realistic fits for the \sot\ ice absorption. 
The laboratory data used to fit the feature at 7.6 ~$\mu$m are provided in Appendix \ref{so2_lab}.

\begin{figure*}[h]
  \centering
  \includegraphics[scale=0.7, trim=0.4cm 4cm 0.4cm 4cm, clip]{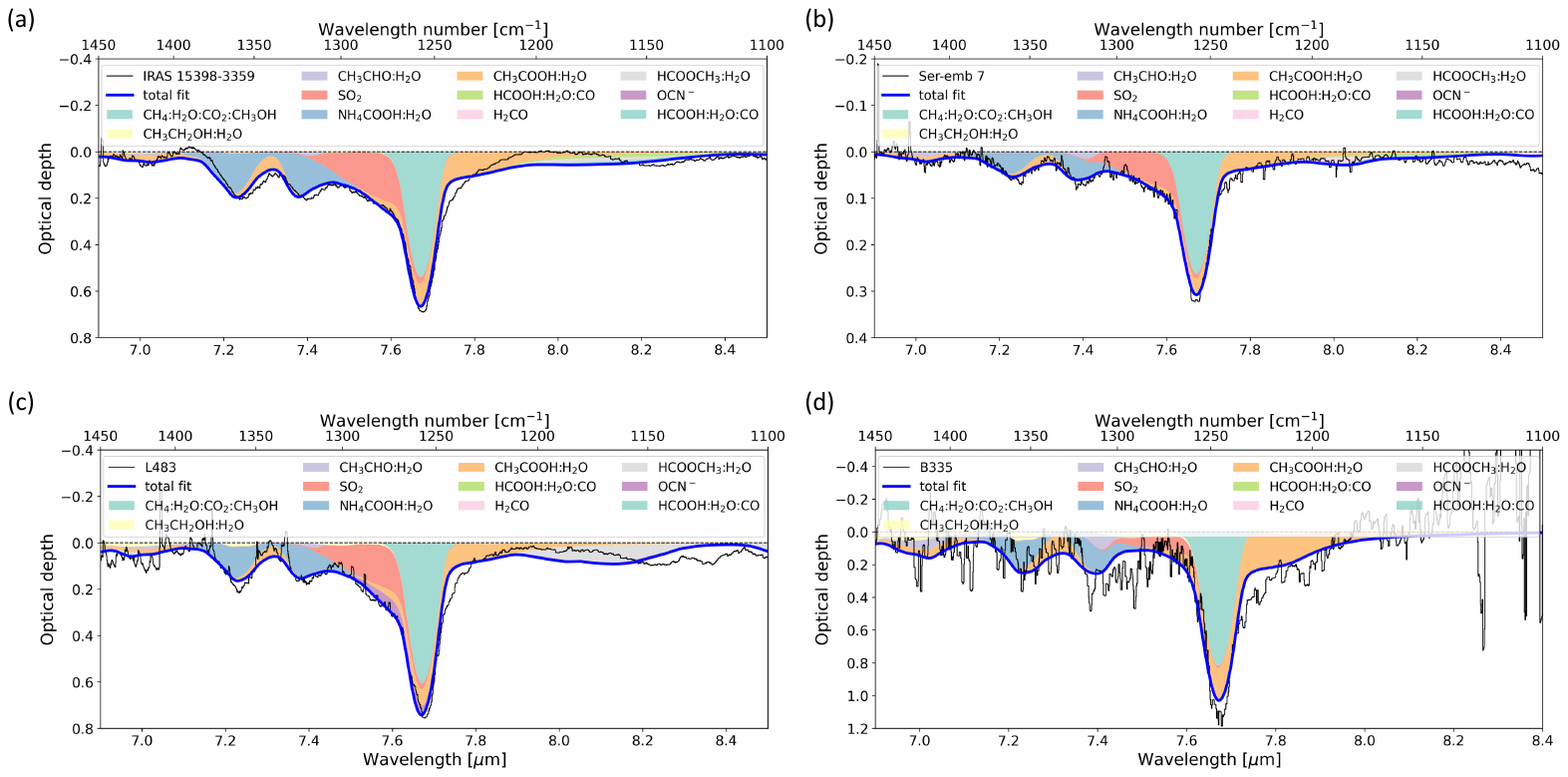} 
  \caption{The \texttt{Omnifit} results for a single iteration for each source with a pure \sot\ mixture. }
  
  \label{omni_sulfur_pure}
\end{figure*}

\noindent {\underline{Pure SO$_2$ fit}}: Using the pure SO$_2$ laboratory spectra, significant SO$_2$ ice is necessary to reproduce the 7.6 $\mu$m shoulder in all four sources (red in Figure \ref{omni_sulfur_pure}). 
The derived SO$_2$ column densities in these cases are of order $10^{17}$cm$^{-2}$ for IRAS 15398-3359 and L483, while the values for Ser-emb 7 and B335 are around $10^{16}$cm$^{-2}$ (Table \ref{ratios}). Ser-emb 7 has an average $N$(SO$_2$)$\sim6.4\times10^{16}$cm$^{-2}$ in the pure SO$_2$ fit, corresponding to an SO$_2$/CH$_4$ ratio of $\sim0.18$. L483 shows a similar ratio of SO$_2$/CH$_4=0.21$. IRAS 15398-3359 exhibits the highest ratio in the pure SO$_2$ case, about 0.27, with $N$(SO$_2$)$\sim1.95\times10^{17}$cm$^{-2}$. In contrast, B335’s spectrum shows lower SO$_2$ column densities in the pure \sot\ ice scenario, where the 7.6~$\mu$m feature showing almost no blue shoulder, and yielding a lower ratio of $\sim0.04$ and a column density of $\sim5\times10^{16}$~cm$^{-2}$. 

Gas-phase SO$_2$ has been reported toward B335, which suggests that an SO$_2$ ice reservoir should also be present \citep[e.g.,][]{Okoda2022_B335}. However, our SO$_2$ ice detections in this source remain tentative given the current signal-to-noise level of the MIRI data, and the column density cannot be constrained with the same confidence as in the other three sources. While the best-fit values tentatively suggest a lower SO$_2$ ice abundance, this difference could reflect the higher noise in the B335 spectra rather than a true abundance variation. Optically thin isotopologue observations (e.g., $^{34}$SO$_2$), which would allow quantitative gas-phase column density estimates, have not yet been obtained for B335. Surveys of other protostars find SO$_2$ gas column densities typically in the range of $\sim$10$^{16}$–10$^{17}$ cm$^{-2}$ \citep{Santos2024b}, but without comparable measurements for this source, a direct gas-to-ice comparison is not possible. If future observations confirm that the SO$_2$ ice abundance is indeed lower than this typical gas-phase range, a plausible explanation could be partial sublimation of SO$_2$-bearing ices during the recent luminosity burst reported by, which would enhance gas-phase SO$_2$ while potentially reducing its solid-phase abundance \citep{Lee2025_B335}. In this scenario, the apparent low SO$_2$ ice abundance may reflect the effects of short-term heating from the recent luminosity burst rather than an intrinsic sulfur deficiency.

\begin{figure*}[ht!]
  \centering
  \includegraphics[scale=0.7, trim=0.4cm 4cm 0.4cm 3cm, clip]{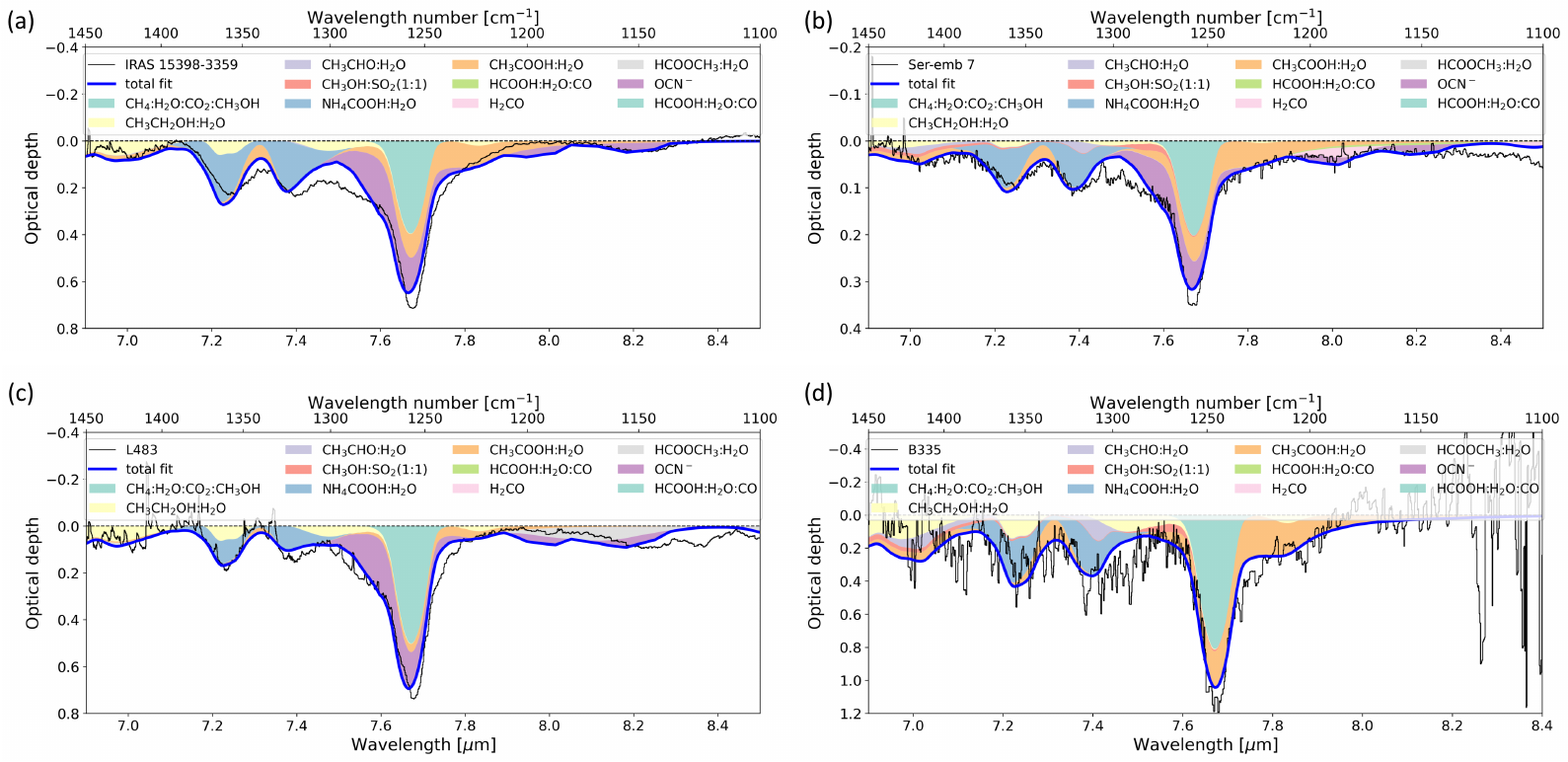} 
  \caption{The \texttt{Omnifit} results for a single iteration for each source with a \sot:\methanol\ mixture. All other laboratory data inputs remain the same as in the pure \sot\ case.}
  
  \label{omni_sulfur_mix}
\end{figure*}

\noindent {\underline{Mixed SO$_2$ with CH$_3$OH fit}}: Using the 1:1 mixed SO$_2$:CH$_3$OH laboratory data,  the contribution of SO$_2$ on the shoulder of the CH$_4$ ice absorption is lower for all sources compared to the fit assuming pure SO$_2$ (Figure~\ref{omni_sulfur_mix}). 

\begin{figure*}[ht!]
  \centering
  \includegraphics[scale=0.7, trim=0.4cm 4cm 0.4cm 3cm, clip]{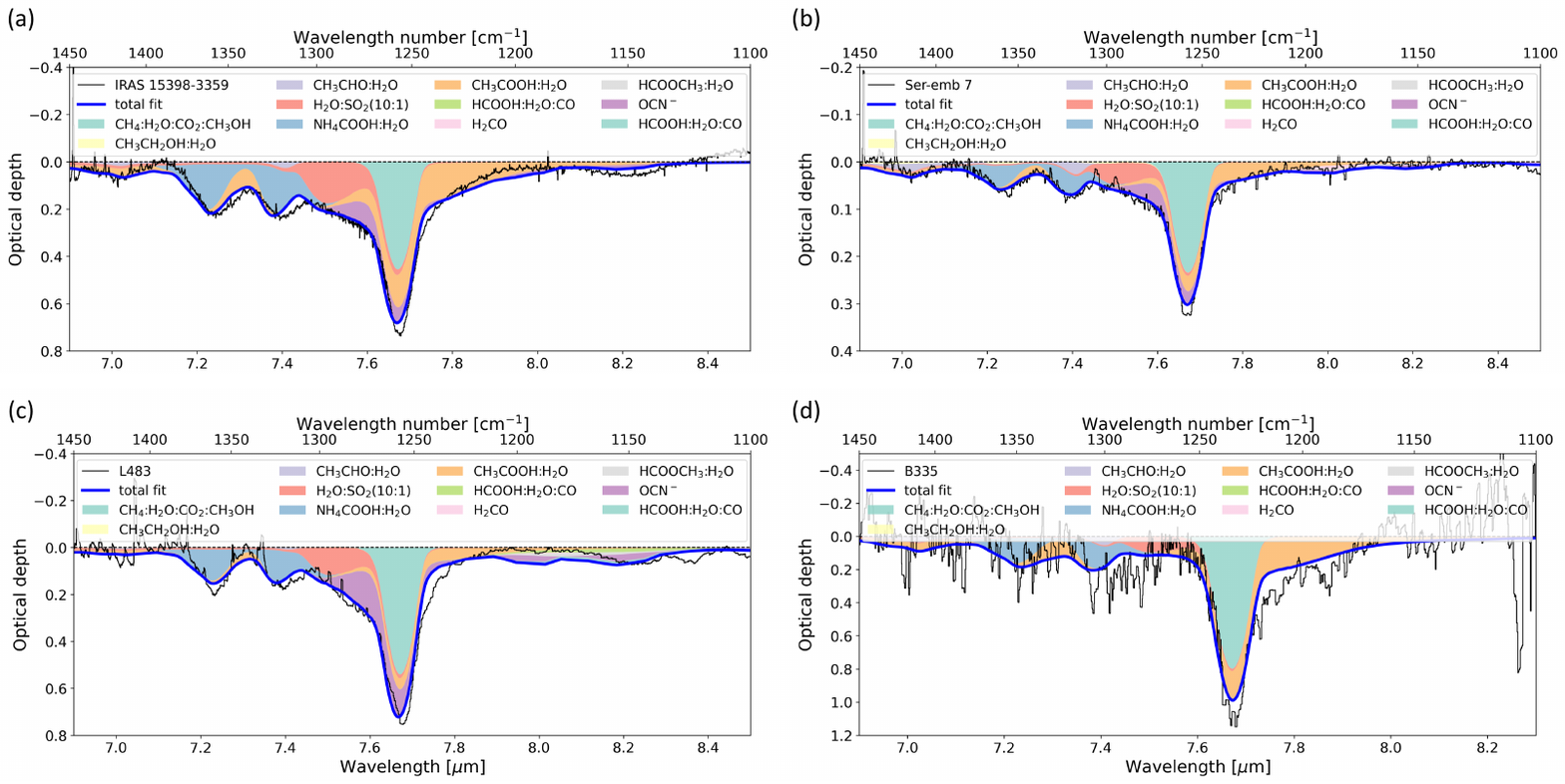} 
  
  \caption{The \texttt{Omnifit} results for a single iteration for each source with a \water:\sot\ (10:1) mixture. All other laboratory data inputs remain the same as in the pure \sot\ case.}
  
  \label{omni_water_mix}
\end{figure*}

When the 1:1 SO$_2$:CH$_3$OH mixture is used, the fitted SO$_2$ contribution decreases across all sources, while the OCN$^-$ component becomes more prominent around 7.3–7.6 $\mu$m. In these fits, the derived OCN$^-$ column density increases by nearly an order of magnitude from about $\sim1\times10^{17}$cm$^{-2}$ in the pure or H$_2$O:SO$_2$ cases to about $\sim1\times10^{18}$cm$^{-2}$ with the SO$_2$:CH$_3$OH mixture. This increase is unlikely to be physical because it exceeds the independently derived NIRSpec OCN$^-$ column densities and occurs alongside a visibly poorer fit to the 7.4–7.6 $\mu$m shoulder (Figure~\ref{omni_sulfur_mix}). The most likely explanation is that this behavior arises from a fitting degeneracy in the modeling routine: because the weaker, broader SO$_2$:CH$_3$OH band overlaps with the OCN$^-$ profile, the algorithm can partially substitute one component for the other to minimize residuals. As a result, the fitting routine compensates for the reduced SO$_2$ contribution by increasing the OCN$^-$ scaling to reproduce the local curvature. We therefore regard the resulting SO$_2$ column densities as lower limits to the SO$_2$/CH$_4$ ratio.

The CH$_4$ feature also appears slightly more blue-shifted than in the pure SO$_2$ case. In Ser-emb 7 and L483, the mixed-profile fits still find a non-zero SO$_2$ contribution, but lower by approximately an order of magnitude than in the pure case (Table \ref{ratios}). For instance, Ser-emb 7 and L483’s SO$_2$/CH$_4$ ratio drops to $\sim0.04$ with the mixed profile. In IRAS 15398-3359, the mixed-profile fit yields a very low SO$_2$ column of $N$SO$_2$ $\sim4\times10^{15}$cm$^{-2}$, resulting in a ratio of SO$_2$/CH$_4$  $\sim0.01$. B335’s SO$_2$/CH$_4$ ratio is similarly small, with the SO$_2$ column density dropping by about a factor of three to $\sim1.4\times10^{16}$cm$^{-2}$. While the 1:1 SO$_2$:CH$_3$OH mixture may still contribute to the CH$_4$ absorption profile, the pure and H$_2$O:SO$_2$ fits provide a better overall representation of the observed spectra.

\noindent {\underline{Mixed SO$_2$ with H$_2$O fit}}: Using the \water:\sot\ (10:1) laboratory data, we find average \sot/\methane\ ratios that fall in between the previous two cases. Ser-emb 7 and L483 have mean \sot/\methane\ ratios of 0.12 and 0.15, respectively, while IRAS 15398-3359 has a slightly higher mean ratio at 0.2, comparable to the fit made using pure \sot\ (0.27). B335 has the lowest \sot/\methane\ ratio of 0.05. Interestingly, we see a combination of the water-mixed \sot\ as well as OCN$^-$ providing the optimal fit for the 7.6~$\mu$m feature for the three sources, IRAS 15398-3359, Ser-emb 7 and L483 (see Figure~\ref{omni_water_mix}).

\begin{table*}[t!]
\caption{SO$_2$/CH$_4$ Ratios and Ice Column Densities}
\label{ratios}
\begin{tabular}{llllcl}
\hline\hline
Source & Scenario & 
Mean {$N_{\mathrm{SO_2}}$} &
Mean {$N_{\mathrm{CH_4}}$} &
Mean ratio (SO$_2$/CH$_4$)$_\mathrm{ice}$ &
{$N_{\mathrm{H_2O}}$(10$^{19}$cm$^{-2}$)} \\
\hline
Ser-emb 7
  & Pure               & 0.64 $\pm$ 0.11  & 3.55 $\pm$ 0.08  & 0.18 $\pm$ 0.03  & 1.11  \\
  & CH$_3$OH:SO$_2$    & 0.108 $\pm$ 0.077 & 2.82 $\pm$ 0.13  & 0.04 $\pm$ 0.03  & \nodata   \\
  & H$_2$O:SO$_2$      & 0.382 $\pm$ 0.071 & 3.31 $\pm$ 0.08  & 0.12 $\pm$ 0.02  & \nodata   \\
  \hline
IRAS 15398-3359
  & Pure               & 1.95 $\pm$ 0.15  & 7.22 $\pm$ 0.09  & 0.27 $\pm$ 0.02  & 1.78 \\
  & CH$_3$OH:SO$_2$    & 0.039 $\pm$ 0.11  & 5.26 $\pm$ 0.23  & 0.01 $\pm$ 0.03  & \nodata    \\
  & H$_2$O:SO$_2$      & 1.32 $\pm$ 0.18  & 6.52 $\pm$ 0.14  & 0.20 $\pm$ 0.03  & \nodata   \\
  \hline
L483
  & Pure               & 1.79 $\pm$ 0.44  & 8.27 $\pm$ 0.35  & 0.21 $\pm$ 0.05  & 2.36 \\
  & CH$_3$OH:SO$_2$    & 0.264 $\pm$ 0.23  & 6.64 $\pm$ 0.23  & 0.04 $\pm$ 0.04  & \nodata   \\
  & H$_2$O:SO$_2$      & 1.07 $\pm$ 0.27  & 7.35 $\pm$ 0.16  & 0.15 $\pm$ 0.04  & \nodata    \\
\hline
B335
  & Pure               & 0.58 $\pm$ 0.27  & 11.0 $\pm$ 1.02   & 0.05 $\pm$ 0.025  & 4.80 \\
  & CH$_3$OH:SO$_2$    & 0.143 $\pm$ 0.63 & 21.6 $\pm$ 0.85   & 0.01 $\pm$ 0.02  & \nodata    \\
  & H$_2$O:SO$_2$      & 0.566 $\pm$ 0.358 & 11.2 $\pm$ 1.082   & 0.05 $\pm$ 0.031  & \nodata    \\
\hline
\end{tabular}
\vspace{2pt}
\begin{minipage}{0.95\textwidth}
\small
\textbf{Notes.} All column densities are in units of 10$^{17}$ molecules cm$^{-2}$, unless noted.
``Pure'' denotes using a pure SO$_2$ laboratory spectrum; ``CH$_3$OH:SO$_2$'' denotes using a CH$_3$OH:SO$_2$ (1:1) lab mixture, and ``H$_2$O:SO$_2$'' denotes using a H$_2$O:SO$_2$ (10:1) lab mixture.
Mean Ratio and column densities for SO$_2$ and CH$_4$ refers to the mean of all 50 baseline iterations. The error is derived from the standard deviation of the baseline uncertainty and the uncertainty in the \texttt{Omnifit} fitting, combined in quadrature for an absolute uncertainty.
All values are derived from the optical depth fits in the 6.8-8.5~$\mu$m region. See Appendix~\ref{water_ice} for fitting of \water\ ice column densities.
\end{minipage}
\end{table*}

\begin{figure*}
  \centering
  \includegraphics[scale=0.5]{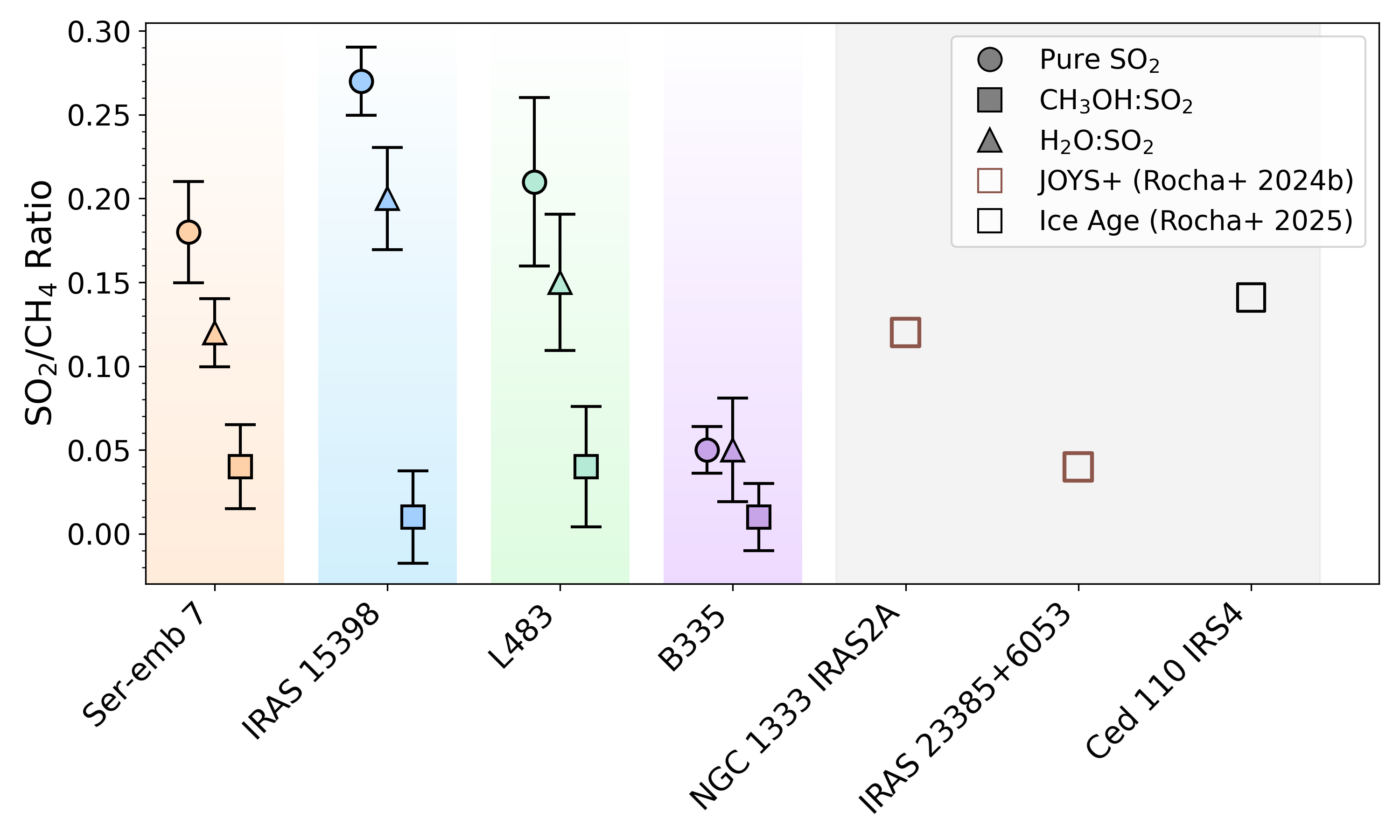}
 
\caption{Average SO$_2$/CH$_4$ column density ratios for the four CORINOS sources. The different markers correspond to the different mixtures used to fit SO$_2$ described in Section~\ref{so2_fits}. The error bars are the absolute uncertainty from both the baseline placement and the \texttt{Omnifit} uncertainty, described in Section~\ref{abs_error}. Despite minor source-to-source variations, the SO$_2$/CH$_4$ ratios are overall similar within a factor of two across the sample.}

\label{so2_methane_comparison}
\end{figure*}

For these three sources, OCN$^-$ contributes column densities of $\sim1.8-7.9\times10^{17}$cm$^{-2}$ for the 7.6 $\mu$m feature. These values are similar to those found in \citet{rocha2024}, where OCN$^-$ was reported at a column density of $\sim1-4\times10^{17}$cm$^{-2}$. The better overall fit to the data in this spectral range, the more realistic ice mixtures used, and the comparability to previous studies leads us to conclude that the fits using the \water:\sot\ (10:1) ice mixture is the most realistic of the three combinations of laboratory data we tested.
The resulting SO$_2$/CH$_4$ ratios for this mixture are: Ser-emb~7: $0.12 \pm 0.02$, IRAS~15398-3359: $0.20\pm 0.03$, L483: $0.15 \pm 0.04$, and B335: $0.05 \pm 0.031$.
These three laboratory cases demonstrate the range of plausible SO$_2$ ice environments in our sample. 
Overall, despite source-to-source differences and baseline-driven uncertainties, the SO$_2$/CH$_4$ ratios are broadly consistent within a factor of two across the four protostars. 
Figure~\ref{so2_methane_comparison} shows the ratios from all three SO$_2$ laboratory ice cases (pure, CH$_3$OH:SO$_2$, H$_2$O:SO$_2$), and Table~\ref{ratios} lists the corresponding values.

\subsection{Impact of Baseline Placement}

The placement of the baseline is a significant factor contributing to systematic uncertainty in ice spectral fitting. This uncertainty is not from random spectral noise, but from the difficulty in determining the true underlying continuum beneath the absorption features. This is particularly relevant in observations characterized by low signal-to-noise ratios or in instances where the continuum is either partially or completely obscured by broad absorption features. Given that the continuum requires estimation rather than direct measurement, varying plausible baseline placements can result in different optical depth spectra, thereby affecting the results of spectral decomposition.

Each iteration in our study produces a slightly different continuum-subtracted optical depth spectrum. We utilize the reduced $\chi^2$ value reported from \texttt{Omnifit}, that includes the consistent noise estimate outlined in the Methodology in Section \ref{labdata}. The resulting spread in reduced $\chi^2$ values reflect how sensitive the total fit is to baseline choice. The value closest to unity is representative of the overall ``best-fit'' iteration.

Figure \ref{best_fit} demonstrates that IRAS 15398-3359 shows minimal variation between the best–fit and the other iterations, whereas B335 shows more substantial variation, reflecting the impact of lower signal-to-noise and weaker features on both baseline placement and the fitting routine. Notably, despite the larger spread in the total ice fits for B335, the standard deviation of the \sot/\methane\ ratio remains relatively small, suggesting that \sot\ is intrinsically weaker than in the other targets, or we are approaching the noise floor for this source and therefore cannot confirm if \sot\ is present with our current data.

\subsection{Absolute Uncertainty Estimation} \label{abs_error}

\begin{figure}
  \centering
  \includegraphics[scale=0.35]{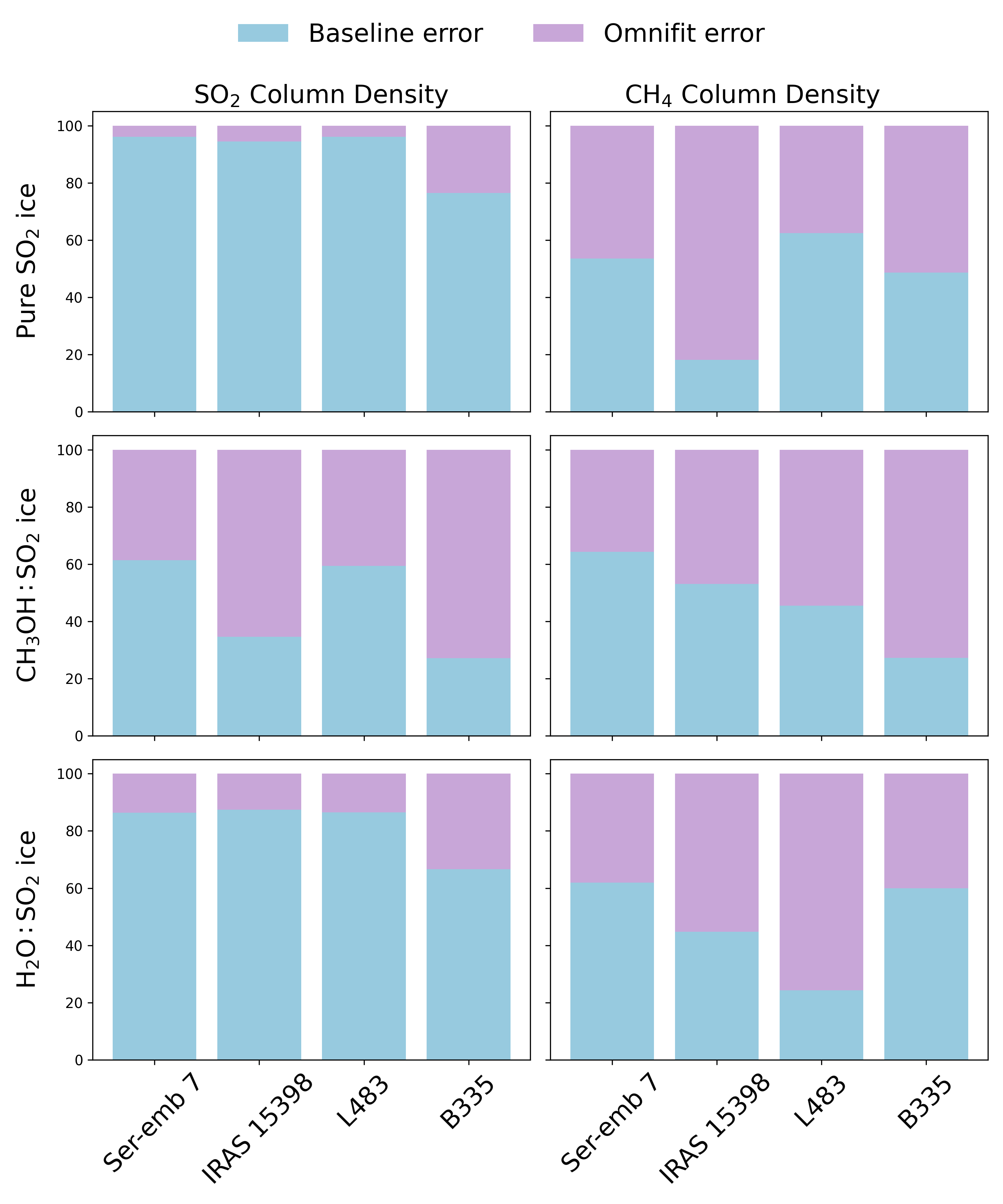}
  
  \caption{Contributions to the total uncertainty on the column density measurements from the placement of baselines (blue) and from \texttt{Omnifit}'s standard deviation from fitting the \sot\ and \methane (purple). Each row shows how the dominant uncertainty on the column density changes for \methane\ and \sot\ for each of the three cases discussed in the text: pure \sot\, \methanol:\sot, and \water:\sot\ laboratory data. 
  } 
  
  \label{uncertainty}
\end{figure}

To get an absolute error in our column density measurements, we combine in quadrature the two independent error terms: (1) the statistical uncertainty that \texttt{Omnifit}'s fitting routine determines from the scalings of the species($\sigma_{\rm fit}$), and (2) the systematic uncertainty induced by random baseline placement ($\sigma_{\rm base}$), which is the standard deviation from our column density calculations for \sot\ and \methane\ for all fifty "iterations" from the \texttt{Omnifit} results. The contributions to the total uncertainty on the column density measurements from each of these sources is illustrated in Figure \ref{uncertainty}.

For example, the pure \sot\ scenario in the top row, the baseline uncertainty (blue) seems to dominate the column densities of \sot\ for all sources, but with the \methane\ column densities, the \texttt{Omnifit} error (purple) contributes more to the overall uncertainty. For the fit using \sot\ mixed with \methanol\, the uncertainty contribution from \texttt{Omnifit} and the choice baseline is closer to equal except for the case of B335. The latter is likely due to an increase in noise in the spectra and the resulting smoothing necessary before fitting. 
The \water:\sot mixture also showed more contribution from baseline uncertainty in the \sot\ column density measurements compared to the the \methane. 

Collectively, we find that the choice of baseline dominates the uncertainty of derived column densities from weaker spectral features (\sot) than than stronger ones (\methane), as expected. However, in all cases, the choice of baseline introduces a non-negligible source of uncertainty in column density measurements compared to the uncertainty introduced by the multi-component ice fitting process alone.

\section{Discussion} \label{discussion}

\subsection{Comparisons to Other Protostellar Regions}\label{other_sources}
\citet{rocha2024} reported SO$_2$ in the Class 0 sources IRAS2A and IRAS23385 using the same SO$_2$/CH$_3$OH (1:1) ice mixture adopted in our analysis, with SO$_2$/CH$_4$ ratios of $\sim0.04–0.12$, comparable to our results under the \methanol:\sot\ ice scenario. Similarly, the JWST IceAge program reported SO$_2$ in the Class I source Ced 110 IRS4A at $\sim1.1\times10^{16}$~cm$^{-2}$, similar to our Ser-emb 7 result using the CH$_3$OH-mixed SO$_2$ ice \citep{Rocha2025}. However, in our spectra, the SO$_2$ mixed with \water\ laboratory data provided a much better fit to the 7.6~$\mu$m shoulder feature in Ser-emb 7, increasing the SO$_2$ column density to $\sim3.8\times10^{16}$~cm$^{-2}$.

While not a direct comparison, the JWST IceAge program also tentatively detected SO$_2$ in dense molecular clouds, with SO$_2$/H$_2$O upper limits of $\sim$0.1–0.3\%, corresponding to column densities of $\sim4\times10^{15}$~cm$^{-2}$. These are 1–2 orders of magnitude lower than our results, though their CH$_4$ column densities are comparable \citep{mcclure2023ices}.

\subsection{Availability of laboratory data}
These results highlight the importance of ice matrix effects previously noted by \citet{Boogert1997}, the profile of the 7.6~$\mu$m SO$_2$ band can vary with ice composition, which can affect how we decompose the blended spectral features in this region. A critical limitation in the interpretation for \sot\ ice for this type of study is the small amount of laboratory SO$_2$ mixtures available for download on the ice spectral databases. As also emphasized by \citet{Rocha2025}, many spectral fits rely on just one or two SO$_2$ combination profiles, primarily SO$_2$:CH$_3$OH (1:1), which may not fully represent the diversity of the \sot\ ice environment. Expanded laboratory studies with SO$_2$ mixed with other major ice constituents across a range of temperatures and mixing ratios would significantly improve the fitting reliability and strengthen confidence in the detection and quantification of \sot\ ice.

\subsection{Missing Sulfur}

The depletion of sulfur in the gas phase by orders of magnitude relative to solar abundances has long been known as the ``missing sulfur'' problem \citep[e.g.,][]{Woods2015,laas2019}. With a solar sulfur-to-hydrogen ratio of [S/H]$_\odot$ = 1.32$\times10^{-5}$ \citep{Asplund2009}, only a small fraction, less than 1\%, is observed in volatile forms in dense molecular clouds \citep{mcclure2023ices}. While solid OCS is a securely identified sulfur-bearing ice \citep{Boogert2022, mcclure2023ices, Taillard2025}, other sulfur ice species like SO$_2$ have been more challenging to detect due to their weak and often blended absorption features. Our new measurements of \sot\ ice in low-mass Class 0 sources provides increased evidence that a non-negligible portion of sulfur resides in the volatile ice reservoir.

\begin{table*}[t]
\centering
\caption{Estimated SO$_2{ice}$/S$_\odot$ from this study}
\label{so2_h_ratios}
\begin{tabular}{lccc}
\hline\hline
Laboratory Ice  
  & $N_{\rm SO_2}/N_{\rm CH_4}$ 
  & $N_{\rm SO_2}/N_{\rm H}$\,($10^{-7}$)$^\dagger$ 
  & SO$_2$/S$_\odot$\,(\%)$^\dagger$ \\
\hline
Pure SO$_2$  
  & 0.18–0.27 
  & 3.6–16.2 
  & 0.3–1.2 \\
CH$_3$OH:SO$_2$ (1:1) 
  & 0.01–0.04 
  & 0.2–2.4 
  & 0.02–0.18 \\
H$_2$O:SO$_2$ (10:1)
  & 0.12–0.20 
  & 2.4–12.0 
  & 0.2–0.9 \\
\hline
\end{tabular}

\vspace{1ex}
\footnotesize
$^\dagger$ 
$N_{\rm SO_2}/N_{\rm H}$ computed as 
$(N_{\rm SO_2}/N_{\rm CH_4})\times(N_{\rm CH_4}/N_{\rm H_2O}=0.01)\times({\rm H_2O/H})$, 
using ${\rm H_2O/H}=2\times10^{-5}$ (lower bound) to $6\times10^{-5}$ (upper bound); 
then SO$_2$/S$_\odot$ $=(N_{\rm SO_2}/N_{\rm H})/(1.32\times10^{-5})\times100\,$\%.
\end{table*}

For the three SO$_2$ scenarios considered in this paper, we estimate the amount of sulfur that could be locked in SO$_2$ ice based on each fit (Table~\ref{so2_h_ratios}). We convert the SO$_2$/CH$_4$ ratios into SO$_2$/H by adopting CH$_4$/H$_2$O=1\% \citep[e.g.,][]{gibb2004,oberg2008,Boogert2015} and a water ice abundance ${\rm H_2O/H}=(2$-$6)\times10^{-5}$ 
consistent with dense cloud observations and protostellar envelopes 
\citep{Boogert2015,vandishoeck2021};
we then express SO$_2$/H as a fraction of the solar sulfur abundance, 
[S/H]$_\odot$ = 1.32$\times10^{-5}$ \citep{Asplund2009}. The exact conversion used is summarized in the notes to Table~\ref{so2_h_ratios}.

Adopting the H$_2$O:SO$_2$ (10:1) laboratory mixture as an illustrative “best-fit” example, the SO$_2$/CH$_4$ ratios translate to ${\rm SO_2/H}\approx(0.24$-$1.20)\times10^{-7}$, corresponding to roughly 0.2-0.9\% of the solar sulfur budget. For the CH$_3$OH:SO$_2$ (1:1) mixture, the implied SO$_2$/H is lower at 0.02-0.18\%, while the pure SO$_2$ case yields 0.3-1.2\%. All ranges are summarized in Table~\ref{so2_h_ratios}.

These levels are also consistent with recent detectability estimates. \citet{Taillard2025} modeled JWST/MIRI thresholds for the 7.6~$\mu$m SO$_2$ feature and found it should be identifiable at SO$_2$/H$_2$O ratios as low as $\sim$0.08\% for a 1~mJy continuum, rising to $\sim$0.5–1.0\% for fainter sources. Our inferred SO$_2$/H$_2$O $\sim$0.2\% lies within that accessible range for deep MIRI observations of embedded protostars.

Even with these detections, the S/H ratio from SO$_2$ ice remains $\sim$10$^{-6}$–10$^{-7}$, still below the cosmic sulfur abundance, and indicating that the majority of sulfur resides in other reservoirs. Recent laboratory and modeling work investigates sulfur locked in more refractory forms, including metal sulfides, sulfur allotropes like S$_8$ \citep{Ferrari2024, laas2019}, or perhaps in ammonium hydrosulfide salts \citep{Slavicinska2025}. \citet{ Slavicinska2025} demonstrated that the 6.85~$\mu$m band which is attributed to NH$_4^+$ can be well fit by NH$_4$SH, or ammonium hydrosulfide. If most NH$_4^+$ is paired with SH$^-$, then NH$_4$SH could account for up to $\sim18$\% of the missing sulfur \citep{Slavicinska2025}. They also report the first tentative detection of SH$^-$ via a weak 5.3 ~$\mu$m feature, further supporting the role of ammonium salts as hidden sulfur reservoirs. These findings reinforce the idea that the missing sulfur is distributed among multiple reservoirs, and our new SO$_2$ measurements add to the inventory of volatile sulfur in the ice phase.

\subsection{SO$_2$ Ice Formation and Chemical Origins}

The detection of SO$_2$ ice in our sample of protostellar envelopes raises important questions about its chemical origins and formation history. Laboratory experiments demonstrate that UV and cosmic-ray irradiation of H$_2$S-rich ices rapidly destroys H$_2$S, efficiently producing oxidized sulfur species such as SO$_2$ and OCS, alongside refractory sulfur compounds \citep{Garozzo2010, Chen2015}. Similarly, UV photons generated by cosmic rays interacting with H$_2$ can photodissociate H$_2$O ice, producing abundant mobile OH radicals even at very low temperatures ($\sim$10–20 K; \citep{Andersson2006, Andersson2008, Ivlev2015, Shingledecker2018}), which can readily oxidize sulfur-bearing species such as S and SO into SO$_2$ \citep{Hama2013, Oba2018, shingledecker2020}. Hydrogen atom bombardment of H$_2$S ice also produces HS and S radicals that can be oxidized into SO$_2$ \citep{Santos2024b}. 

These pathways may operate across multiple evolutionary stages: during the prestellar phase, within collapsing envelopes, or even persisting into the earliest protostellar stages. Observations of gas-phase SO$_2$ and OCS in protostellar sources indicate that their abundances are consistent with an origin in early ice chemistry followed by desorption \citep{Santos2024b}. At the same time, our observed SO$_2$ ice column densities ($\sim10^{16}-10^{17}$~cm$^{-2}$) are consistent with ongoing production within cold protostellar envelopes. Taken together, these results suggest that the presence of SO$_2$ ice likely reflects contributions from both early and ongoing grain-surface chemistry, and does not indicate a major chemical reset. The low SO$_2$ abundances detected toward quiescent dense-cloud sightlines \citep[e.g.,][]{mcclure2023ices} further support this interpretation, demonstrating that SO$_2$ formation is inefficient under cold, weak-UV conditions and increases in response to the onset of protostellar processing.

\citet{Santos2024b} emphasize that viable SO$_2$ formation pathways require a water-rich ice matrix, implying that SO$_2$ ice formation likely occurs early, alongside H$_2$O ice and before significant CO freeze-out, when oxygen is abundant in water rather than locked in CO. Our observed SO$_2$ column densities ($\sim10^{16}-10^{17}$~cm$^{-2}$) suggest that a small yet significant fraction of sulfur remains trapped in these oxygen-rich ice forms. The presence of SO$_2$ ice may therefore serve as a valuable observational tracer of early-stage grain-surface oxygen chemistry, placing constraints on the timing and efficiency of sulfur incorporation into protostellar ice mantles.

These findings highlight the importance of studying Class 0 protostars, which represent the earliest evolutionary stage accessible for tracing volatile sulfur chemistry before it becomes more challenging to observe in later evolutionary phases. As protostars evolve toward Class II disk systems, increased UV irradiation, thermal desorption, and chemical reprocessing can obscure detections of simpler sulfur species such as SO$_2$ and SO. Nonetheless, strong detections of multiple SO$_2$ and SO transitions have been reported in certain disks, such as Oph IRS 48, and SO has been identified in additional disk surveys \citep{Booth2021, Booth2024a, Booth2024b}. Surveys of embedded Class 0/I sources, such as PEACHES, frequently detect SO$_2$ gas, reinforcing its presence in cold, early protostellar environments and supporting the interpretation that sulfur-bearing ices like SO$_2$ predominantly form during the earliest stages of star formation \citep{ArturdelaVillarmois2023}. For quantitative comparison with ice abundances, however, isotopologues such as $^{34}$SO$_2$ are required, since the main SO$_2$ lines can be optically thick \citep[e.g.,][]{Booth2024b}. 
For context, gas-phase $^{34}$SO$_2$ observations toward the Class 0 source IRAS 16293-2422 yield SO$_2$ column densities of order 10$^{16}$ cm$^{-2}$ \citep{Drozdovskaya2018}, about an order of magnitude lower than our solid-state SO$_2$ estimate for IRAS 15398-3359, suggesting that a substantial fraction of sulfur may remain in the ice at this stage.

While detailed analysis requires optically thin isotopologues, a tentative comparison of SO$_2$/CH$_3$OH ratios in the gas and ice is still informative. In hot corinos and cores, gas-phase SO$_2$/CH$_3$OH ratios are typically $\sim$10$^{-4}$–10$^{-3}$ (e.g., 1.3$\times$10$^{-4}$ in IRAS~16293B and 1.3$\times$10$^{-3}$ in G31.41; \citet{Belloche2025}). Using our SO$_2$/CH$_4$ average ratios of 0.1–0.2, combined with typical CH$_4$/CH$_3$OH(ice) values of 0.2–0.5 for Class 0/I protostars \citep{oberg2008, Boogert2015}, corresponds to SO$_2$/CH$_3$OH(ice) ratios of $\sim$0.02–0.10. The relative ratio is $\sim$10$^{2}$–10$^{3}$ times higher in the ice than in the hot gas, indicating that the observed SO$_2$ can be fully supplied by release from ices (e.g., sublimation or sputtering), and that its lower gas-phase abundance likely reflects chemical processing during and after release.

\begin{figure*}
  \centering
  \includegraphics[scale=0.43]{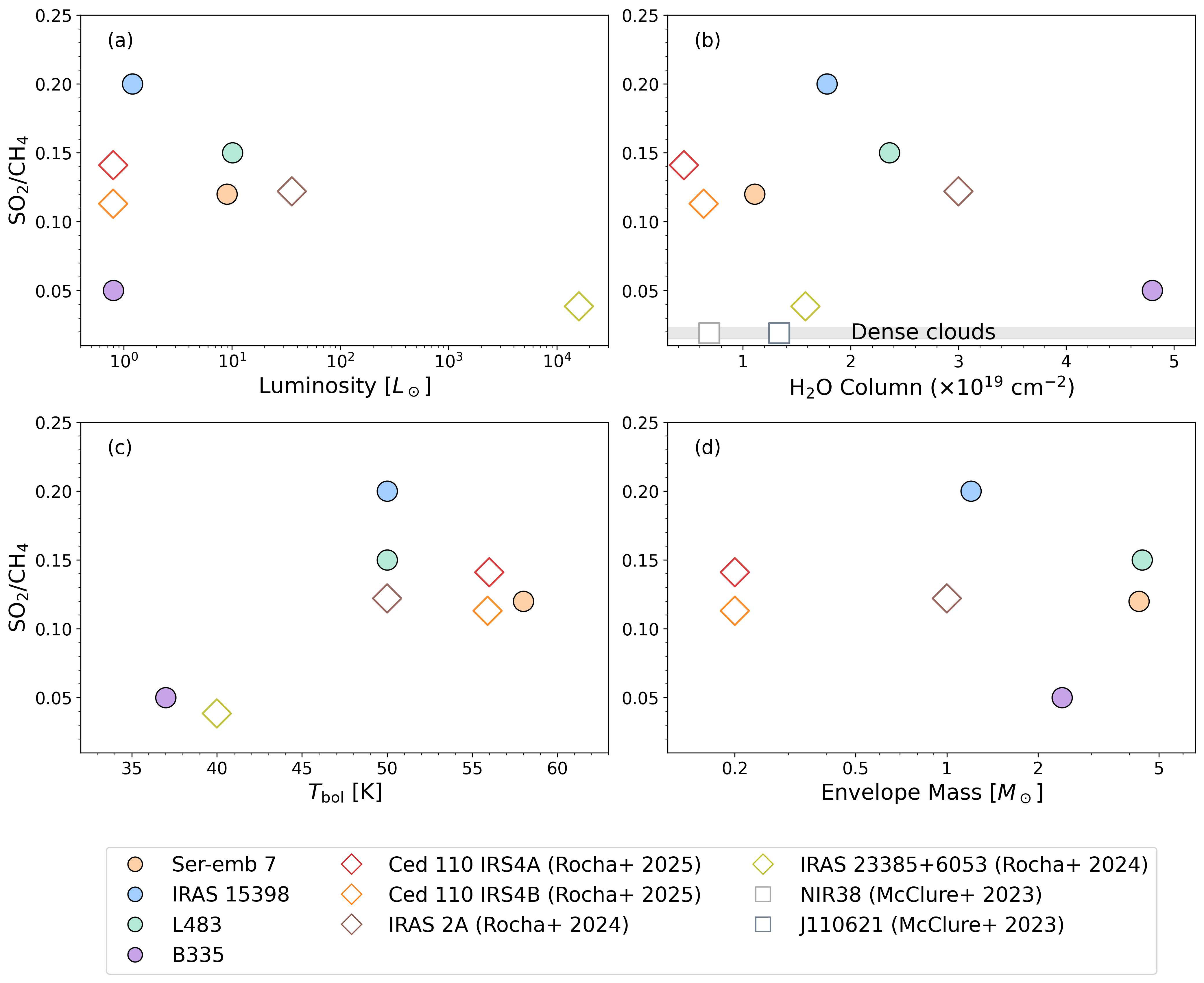}

\caption{(\textbf{a}) SO$_2$/CH$_4$ ratio versus bolometric luminosity for the CORINOS sources and the comparison protostars discussed in Section~\ref{other_sources} (Ced~110 IRS4A/B, IRAS~2A, and IRAS~23385+6053).  
(\textbf{b}) SO$_2$/CH$_4$ ratio versus H$_2$O ice column density. Dense cloud sightlines from the IceAge program \protect\citep[][NIR38 and J110621]{mcclure2023ices} are shown as gray squares, and the shaded region marks the SO$_2$/CH$_4$ range observed in molecular cloud ices.  
(\textbf{c}) SO$_2$/CH$_4$ ratio versus bolometric temperature ($T_{\rm bol}$) for both the CORINOS targets and the comparison protostars. $T_\mathrm{bol}$ for IRAS~2A and Ced~110 IRS4A/B are taken from \citet{Lee2015}; the value for IRAS 23385+6053 is from \citet{Francis2024}.
(\textbf{d}) SO$_2$/CH$_4$ ratio versus envelope mass. Envelope masses for Ced~110 IRS4A/B are taken from \citet{SanJose-Garcia2013}, and the envelope mass for IRAS~2A is taken from \citet{jorgensen2004}. IRAS~23385+6053 is excluded from this panel because its envelope mass is not reliably constrained.  
The CORINOS values use the mean SO$_2$/CH$_4$ ratios from Table~\ref{results}, derived from the H$_2$O:SO$_2$ laboratory fits.
}
\label{luminosity_water_ice}

\end{figure*}

\subsection{Evaluating Environmental Factors on the \sot/\methane\ Ratio}

One goal of the CORINOS program is to determine which environmental factors regulate ice-phase sulfur abundances in Class~0 protostars. We examine the roles of protostellar luminosity, envelope mass, water-ice column density, and source inclination using our four CORINOS targets together with the additional protostars discussed in Section~\ref{other_sources} (Ced~110 IRS4A/B, IRAS~2A, and IRAS~23385+6053). Although the combined sample remains small, a comparison allows us to evaluate whether any broad trends emerge in the SO$_2$/CH$_4$ ratio.

Protostellar luminosity is often invoked as a driver of ice processing, with the strongest effects expected for highly volatile species such as CO \citep{boogert2008, oberg2011sp}. CH$_4$ is less sensitive to heating because it is frequently trapped in H$_2$O-rich matrices, while the temperature-dependent behavior of SO$_2$ in ices is less well constrained. Figure~\ref{luminosity_water_ice}a shows how SO$_2$/CH$4$ varies with bolometric luminosity ($L\mathrm{bol}$) across both the CORINOS and comparison sources. No systematic trend is seen. B335 and IRAS~15398--3359, the two least luminous objects, lie at opposite extremes of the observed ratio range, while the more luminous Ser-emb~7 and L483 cluster at intermediate values. B335 is a known outlier due to its prior luminosity burst \citep{Lee2025_B335}, and its SO$_2$ absorption measurement is highly uncertain because of spectral noise. These comparisons indicate that luminosity alone does not control the SO$_2$/CH$_4$ ratio, consistent with \citet{Santos2024b}, who reported no link between luminosity and SO$_2$/CH$_3$OH abundance ratios. Our measurements therefore contribute several new jointly measured SO$_2$ and CH$_4$ ice abundances to the literature, although the sample is still too small to identify luminosity-dependent behavior.

We next examine whether water-ice column density correlates with the SO$_2$/CH$_4$ ratio (Figure~\ref{luminosity_water_ice}b). Across both the CORINOS and comparison protostars, no clear trend is present. B335, which has the highest H$_2$O column in our sample, shows one of the lowest SO$_2$/CH$_4$ ratios, though this value is highly uncertain due to spectral noise. Ser-emb~7 has the lowest water column among the CORINOS targets yet exhibits an intermediate SO$_2$/CH$_4$ ratio. The comparison protostars show similar scatter: IRAS~2A and Ced~110 IRS4A/B span a modest range of H$_2$O columns while occupying overlapping SO$_2$/CH$_4$ values, and IRAS~23385+6053 lies at the low end of both H$_2$O column and SO$_2$/CH$_4$. Dense cloud sightlines (NIR38 and J110621) appear in the shaded region at the bottom of the panel, where SO$_2$/CH$_4$ values are an order of magnitude lower than in protostars despite comparable or higher water columns. These comparisons show that total H$_2$O ice column density does not set the SO$_2$ abundance. Instead, SO$_2$ production likely depends on the availability of chemical precursors such as atomic S and OH and on early prestellar ice chemistry rather than on the bulk amount of water ice \citep{Santos2024b, Rocha2025}.

$T_\mathrm{bol}$ is a standard indicator of protostellar evolutionary stage, defined as the temperature of a blackbody with the same mean frequency as the observed SED \citep{Myers1993}. Values below roughly 70~K correspond to the deeply embedded Class~0 phase \citep{Chen1995}. When we compare SO$_2$/CH$4$ with $T\mathrm{bol}$ for the CORINOS and comparison sources (Figure~\ref{luminosity_water_ice}c), the ratios remain clustered across the Class~0 range (35--70~K). The coldest objects, including IRAS~23385+6053 and B335, show the lowest ratios, while the slightly warmer Class~0 protostars occupy intermediate values. Although the sample is limited, this pattern suggests that the very earliest, coldest stages may be relatively SO$_2$-poor compared to CH$_4$.

We also examined whether the SO$_2$/CH$_4$ ratio correlates with envelope mass (Figure~\ref{luminosity_water_ice}d). Across the CORINOS sources and the comparison protostars introduced in Section~\ref{other_sources}, the ratios occupy a similar range over more than an order of magnitude in envelope mass. The lowest SO$_2$/CH$_4$ values occur toward IRAS~23385+6053 ($M{\rm env} \approx 0.2~M\odot$; \citealt{rocha2024}) and B335 ($M{\rm env} \approx 2.4~M\odot$; this work) despite their very different envelope masses. Ser-emb~7 and L483, which host the most massive envelopes in the sample (4-5~$M_\odot$; \citealt{jorgensen2004}), show intermediate ratios similar to lower-mass systems. Ced~110 IRS4A and IRS4B, which also host low-mass envelopes ($\sim$0.2~$M_\odot$; \citealt{SanJose-Garcia2013}), fall within the same overall distribution. This pattern indicates that bulk envelope mass is not the dominant factor controlling SO$_2$ ice abundance and instead points toward chemical conditions inherited from prestellar or very early protostellar stages.

Lastly, we explored whether source inclination influences the SO$_2$/CH$_4$ ratio. Our sample spans a broad range of viewing geometries, from nearly face-on (Ser-emb~7, $i \sim 25^\circ$) to moderately inclined (L483 and IRAS~15398--3359, $i \sim 70^\circ$). L483 is a confirmed binary system \citep{Cox2022}, and although its envelope shows a moderately inclined rotation pattern \citep{Jacobsen2019, Hirota2025}, the inclinations of the individual circumstellar disks are unknown. This unresolved structure introduces additional uncertainty when interpreting inclination-dependent behavior. More edge-on sightlines are expected to probe denser envelope and midplane material, enhancing ice absorption \citep{Pontoppidan2005, Crapsi2008, Tobin2008}. JWST studies likewise indicate that moderately inclined sightlines may underestimate ice columns by missing the regions where most ices reside \citep{Rocha2025}, and radiative transfer models show that CO, CO$_2$, and H$_2$O absorption often originates at radii of order $\sim$1000~au (Thompson et al., in prep.). However, within our sample, the SO$_2$/CH$_4$ ratio does not vary systematically with inclination. Geometry therefore influences total ice column but does not appear to regulate relative SO$_2$ abundances.

Overall, no clear correlations are found between SO$_2$/CH$_4$ and protostellar luminosity, envelope mass, water column density, or inclination. The ratios are broadly similar across diverse physical environments, suggesting that macroscopic source properties do not strongly regulate SO$_2$ ice abundance. A more likely explanation is that the observed ratios reflect underlying chemical conditions such as the prestellar sulfur reservoir, grain-surface chemistry prior to collapse, or variations in cosmic-ray ionization rate rather than later-stage protostellar evolution or geometry \citep{Vastel2018, Bulut2021}. This interpretation is supported by recent JWST observations of Ced~110 IRS4, which show that ice-phase chemistry changes little after initial formation in the pre-collapse stage \citep{Rocha2025}. A larger sample will be required to confirm these conclusions.

\subsection{OCN$^-$ and Other Species in the 6.8–8.5~$\mu$m Region}
\label{ocn_discussion}

Although our primary focus was the measurement of SO$_2$ and CH$_4$ ice abundances, our spectral fits included a total of ten ice species to accurately reproduce the blended absorption features observed between 6.8 and 8.5~$\mu$m. Among these, the cyanate ion (OCN$^-$) plays an especially important role. As shown in Section~\ref{ocn_results}, OCN$^-$ is independently detected with NIRSpec, and its measured column densities are consistent with those derived from our MIRI fits in at least one source. This agreement strengthens confidence in its presence and highlights the value of near-infrared constraints for interpreting blended mid-infrared bands. 

Consistent with \citet{rocha2024}, who demonstrated that the 7.68~$\mu$m absorption feature arises from a blend of SO$_2$ and OCN$^-$, we find that OCN$^-$ contributes significantly in several of our sources. In Ser-emb~7 and L483, OCN$^-$ appears across both pure and mixed SO$_2$ fits, while in IRAS~15398$-$3359 it is particularly important when adopting the CH$_3$OH:SO$_2$ mixture. The increased preference for OCN$^-$ in cases where the SO$_2$ profile is broadened or weakened suggests that \texttt{Omnifit} may redistribute absorption strength between the two species. 

Beyond OCN$^-$, ethanol (CH$_3$CH$_2$OH) and acetic acid (CH$_3$COOH) were also fit with relatively consistent appearances across several sources, though definitive detections remain uncertain. Ethanol was most consistently identified in IRAS~15398$-$3359 and L483, yielding average column densities of $8.3\times10^{17}$\,cm$^{-2}$ and $9.8\times10^{17}$\,cm$^{-2}$, respectively, with relative uncertainties of 42\% and 45\%. Acetic acid appeared across all four sources with typical column densities of $\sim3\times10^{17}$\,cm$^{-2}$. Both values are slightly higher than but broadly consistent with those reported by \citet{rocha2024}. 

B335 yielded a significantly higher ethanol column density ($1.2\times10^{19}$\,cm$^{-2}$), but the fit was associated with high uncertainty in continuum placement. Likewise, ethanol detection in Ser-emb~7 exhibited extreme sensitivity to baseline choice, with relative errors exceeding 400\%. Given the relatively weak and blended nature of these features, and their strong dependence on baseline selection, we do not interpret these findings as definitive detections in this study. Rather, their inclusion was essential for achieving sound fits to the primary SO$_2$ and CH$_4$ absorptions, and they should be targeted for future dedicated analysis.

\section{Summary and Conclusions}
\label{summary}

We have presented a new analysis of the 6.8-8.5 ~$\mu$m region in four Class 0 protostars (B335, L483, IRAS 15398-3359, and Ser-emb 7) from the CORINOS JWST MIRI MRS program. This region contains blended ice absorption features from CH$_4$, SO$_2$, and other organic species, and we focus on assessing the impact of baseline uncertainty on column density measurements of \sot/\methane. Our conclusions are as follows:

\begin{itemize}

\item Measurements of CH$_4$ in all four targets show a clear \methane\ band at $\sim$7.7 ~$\mu$m, with column densities on the order of $10^{17-18}$cm$^{-2}$. These values imply that \methane\ is consistently detected in our sample and is consistent with previous ice fittings of low-mass protostars \citep[e.g.,][]{boogert2008}.

\item We show that \sot\ contributes to the blue shoulder of the 7.7 $\mu$m \methane\ band for IRAS 15398-3359, L483, and Ser-emb 7, with column densities on the order of $10^{16}$–$10^{17}$ cm$^{-2}$ using pure SO$_2$ and \water:\sot\ mixed laboratory ice data. For B335, however, the spectra are too noisy to confirm a detection, and any possible contribution from SO$_2$ should be regarded as tentative. This is consistent with \citep{rocha2024}, who reported solid SO$_2$ detections in IRAS 2A but only tentative evidence in IRAS 23385, at similar column densities to our B335 limits. For the sources where SO$_2$ is detected (Ser-emb 7, L483, and IRAS 15398), the inferred SO$_2$/CH$_4$ ratios are 0.18–0.27, with lower values of $\sim$0.04 in B335. When SO$_2$:CH$_3$OH (1:1) mixtures are used instead, the SO$_2$/\methane\ ratios for all sources are reduced by a factor of several, to $\sim$0.01–0.04. Fits with \water:\sot\ laboratory ice data yield intermediate ratios of 0.12–0.20, lying between the pure and \methanol\ mixed cases.

\item OCN$^-$ has a vibrational mode at 7.6~$\mu$m that coincides with the CH$_4$ and SO$_2$ features. This band was present in several of our fits, and the preference for OCN$^-$ increased when the SO$_2$ profile was broadened or weakened, particularly for the pure or CH$_3$OH:SO$_2$ mixtures. By contrast, the H$_2$O:SO$_2$ mixture provided a better simultaneous fit to both SO$_2$ and OCN$^-$. In our sample, OCN$^-$ was consistently detected in Ser-emb~7 and L483 across both pure and mixed SO$_2$ fits, and in IRAS~15398$-$3359 when using the CH$_3$OH:SO$_2$ mixture. Importantly, our independent JWST NIRSpec detection of the 4.6~$\mu$m OCN$^-$ band in IRAS~15398$-$3359 yielded a column density broadly consistent with the value derived at 7.6~$\mu$m when using the H$_2$O:SO$_2$ mixture. This agreement strengthens confidence that both SO$_2$ and OCN$^-$ contribute to the 7.6~$\mu$m absorption, and further supports the conclusion that H$_2$O:SO$_2$ mixtures provide the most realistic fits to the SO$_2$ feature.

\item Our analysis demonstrates that the derived SO$_2$ column densities can vary based solely on the placement of the local continuum baseline. This baseline uncertainty typically dominates over the statistical fitting error from \texttt{Omnifit}, especially for pure SO$_2$ and H$_2$O:SO$_2$ fits. In contrast, the CH$_4$ column densities are relatively insensitive to baseline changes. Consequently, variations in the SO$_2$/CH$_4$ ratio are primarily driven by baseline placement. Nevertheless, across the four sources the ratios remain broadly similar within a factor of two, underscoring both the need for careful baseline treatment and the stability of the overall SO$_2$/CH$_4$ abundance relationship.

\item We find no clear correlation between the SO$_2$/CH$_4$ ratio and luminosity, water-ice column density, envelope mass, or inclination. A slight decrease in SO$_2$/CH$_4$ with lower $T_{\rm bol}$ may be present, but the sample is too small to confirm this. These results suggest that other environmental factors not explored in this work, or initial chemical conditions, could play a key role in regulating sulfur ice chemistry.

\item Finally, we emphasize the small sample of laboratory ice data available for \sot\ ices. Most existing analyses rely on just a few available profiles, such as SO$_2$:CH$_3$OH (1:1), which may not reflect the full diversity of interstellar ice matrices. Additional mixtures are needed to reduce degeneracy in fitting and to more accurately interpret blended absorption features. More laboratory data made available from different temperatures and mixing ratios would significantly improve the ability to identify and quantify \sot\ ice contributions.

\end{itemize}

Our randomized baseline approach reveals how even small changes in continuum placement can alter the measured abundance of a weak feature in the 6.8-8.5 ~$\mu$m region within our four Class 0 sources. The results provide new constraints on the volatile sulfur budget and suggest that \sot\ ice could be responsible for a larger fraction of missing sulfur than previously predicted. However, the sensitivity of the \sot\ column density to baseline choice emphasizes the need for caution when interpreting these weak features, as well as the importance of statistical fitting methods.

\section{Acknowledgments}

This work is based on observations made with the NASA/
ESA/CSA James Webb Space Telescope. The data were obtained from the Mikulski Archive for Space Telescopes at the Space Telescope Science Institute, which is operated by the Association of Universities for Research in Astronomy, Inc., under NASA contract NAS 5-03127 for JWST. These observations are associated with JWST GO Cycle 1 program ID 2151. The JWST data used in this paper can be found in MAST. R.E.G., L.I.C. and JDG acknowledge funding from JWST-GO-02151.001-A. R.E.G. acknowledges support from the UVA Interdisciplinary Doctoral Fellowship. Y.-L.Y. acknowledges support from Grant-in-Aid from the Ministry of Education, Culture, Sports, Science, and Technology of Japan (20H05844 and 25H00676). L.I.C. also acknowledges support from the Research Corporation for Science Advancement Cottrell Scholarship Award 28249, the David and Lucille Packard Foundation, and NSF AAG Award 2205698. R.T.G. thanks the National Science Foundation for funding through the Astronomy \& Astrophysics program (grant number 2206516). R.E.G. thanks Dr. Suchitra Narayanan (Center for Astrophysics | Harvard \& Smithsonian)) for performing preliminary laboratory experiments that helped guide the direction of this work. We also thank the anonymous referee for constructive feedback that improved the clarity of this manuscript.

\vspace{5mm}
\facilities{JWST}

\software{\texttt{Omnifit} \citep{Omnifit2015}}


\appendix
\twocolumngrid

\section{\sot\ Laboratory Data} \label{so2_lab}

\begin{figure}[h]
  \centering
  \includegraphics[scale=0.2]{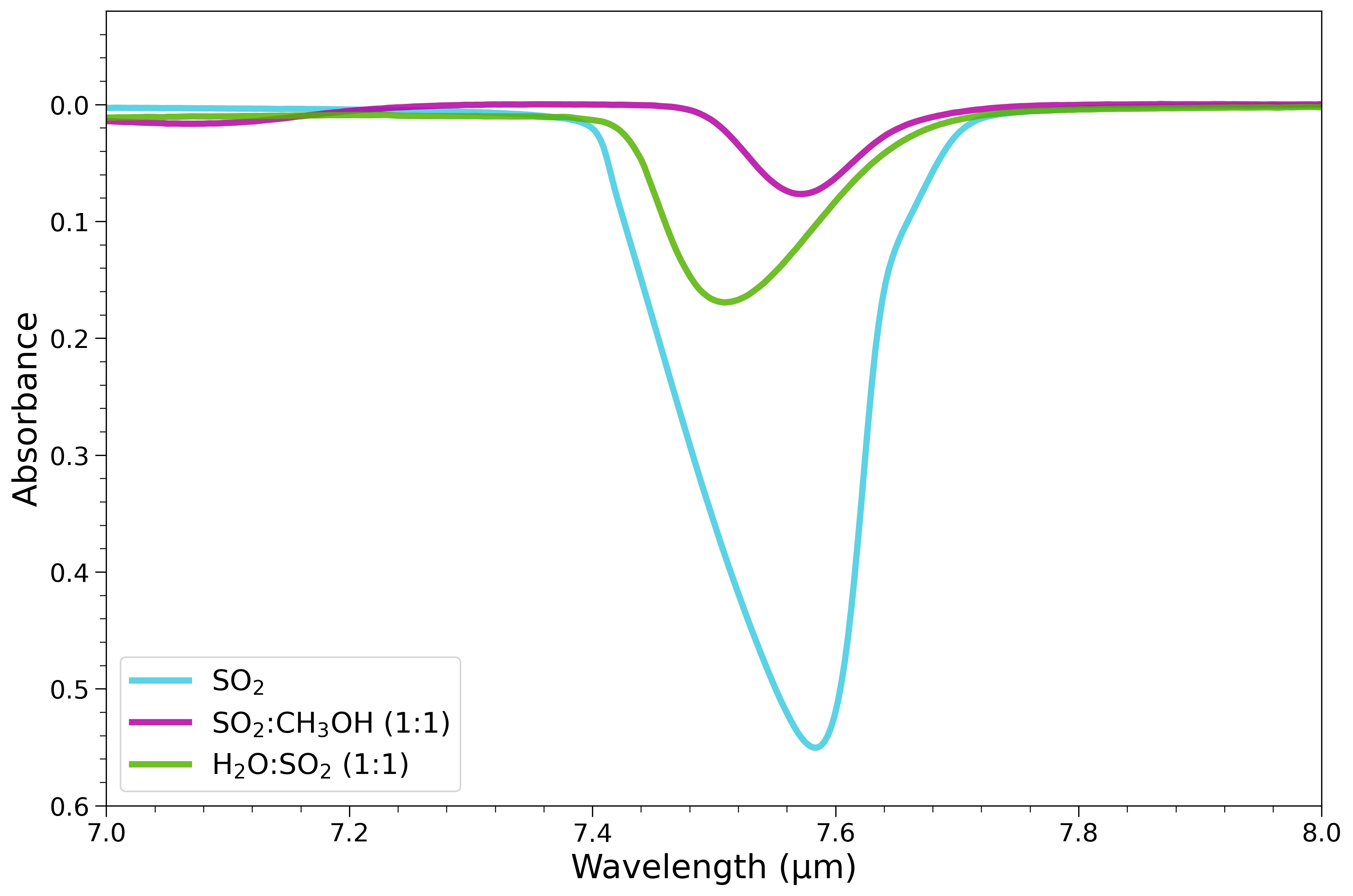}
  
  \caption{\sot\ laboratory data for pure and 
\methanol\ mixed ices taken from \cite{Boogert1997} and the Goddard Ice Database. Note how the peak of \sot\ changes depending on mixture and temperature. Only the low temperature pure and mixed ice were used in this study for comparison. }
  
  \label{so2_compare}
\end{figure}

The differences between the pure \sot, \water:\sot\ (10:1), and \sot:\methanol (1:1) laboratory spectra are due to the mixed matrix altering the vibrational mode for the pure species. In the mixture, \methanol\ causes broadening and weakening of the 7.6 ~$\mu$m absorption feature. Despite unknown thickness in the mixed \methanol\ ice experiment, the use of \texttt{Omnifit} compensates for such differences by scaling the laboratory profiles during fitting. The reduced contribution of the mixed \sot\ spectrum in the \texttt{Omnifit} results is not due to the absolute strength of the feature seen in the lab data of Fig. \ref{so2_compare}, but because the broader, shallower profile does not match as well to the observed absorption feature compared to the pure \sot\ profile \citet{Boogert1997}, or the \water:\sot\ profile \citet{Yarnall2022}. The pure and \methanol\ mixed ices are the only available \sot\ ice data on the LIDA database to download and use in the fittings of this feature, which stresses the necessity of obtaining more laboratory data for \sot. The \water:\sot\ spectra is taken from \citet{Yarnall2022}, and the feature is blue-shifted compared to the \methanol\ mixture.

\section{H$_2$O ice measurements} \label{water_ice}
Water ice is the most abundant species in ice mantle. While our MIRI MRS spectra do not cover the strongest \water\ stretching mode at $\sim3$ \micron, we can use the \water\ libration mode at $\sim$13.6 \micron\ to measure the \water\ ice column density. The broad libration mode mostly blends with the silicate dust absorption at $\sim10$ \micron. To model this \water\ band, we use the silicate-subtracted optical depth spectra presented in Turner et al. (in prep.). We describe a brief outline of the steps leading to these spectra, while refer the details to their study. The silicate dust absorption is fitted with two compositions, pyroxene and forsterite, simultaneously with a polynomial baseline. The anchor points are chosen at $\sim5$, $\sim8$ and $\gtrsim20$ \micron. Figure \ref{water_ice} shows the silicate-subtracted optical depth spectra of the CORINOS sources. The spectrum of L483 is further smoothed with a median filter using a kernel size of 31 to remove the impact of emission lines. Moreover, we removed residual offset in the optical spectrum of L483 with a simple straight baseline (Figure\,\ref{water_ice}).

Figure\,\ref{water_ice} shows clear absorption feature at 10-18 \micron, consistent with the \water\ ice libration mode. We started the modeling with only a low-temperature \water\ ice component (e.g., 15 K); however, the broad absorption feature at 10-18 \micron\ cannot be reproduced with a single component. We employed a low- and high-temperature \water\ component, 15 K and 135 K, to fit the feature. The \water\ ice laboratory spectra are taken from \citet[]{Oberg2007icemixtures}, processed with a local baseline subtraction fitted with a second-order polynomial to isolate the libration mode. We optimized the contributions from these two \water\ ice components using the Trust Region Reflective algorithm and least squares minimization implemented in \texttt{astropy} \citep{astropy:2013,astropy:2018,astropy:2022}.  Figure \ref{water_ice} shows the best-fitting \water\ ice spectra and Table \ref{N_h2o} lists the fitted \water\ column densities.

\begin{table}[htbp!]
    \centering
    \caption{Fitted \water\ column densities}
    \begin{tabular}{c|ccc}
        \toprule
        Source              & 15 K          & 135 K         & Total         \\
                            & \multicolumn{3}{c}{(10$^{19}$ cm$^{-2}$)} \\
        \midrule
        B335                & 4.80   & \nodata & 4.80   \\
        IRAS 15398$-$3359   & 0.27   & 1.51   & 1.78   \\
        L483                & 1.43   & 0.93   & 2.36   \\
        Ser-emb 7           & 1.11   & \nodata & 1.11   \\
        \bottomrule
        \multicolumn{4}{l}{%
            \parbox[t]{0.9\linewidth}{%
                Note: The high-temperature \water\ ice component is not detected in B335 and Ser-emb 7.%
            }%
        }\\
    \end{tabular}
    \label{N_h2o}
\end{table}

\begin{figure*}
    \centering
    \includegraphics[width=0.48\linewidth]{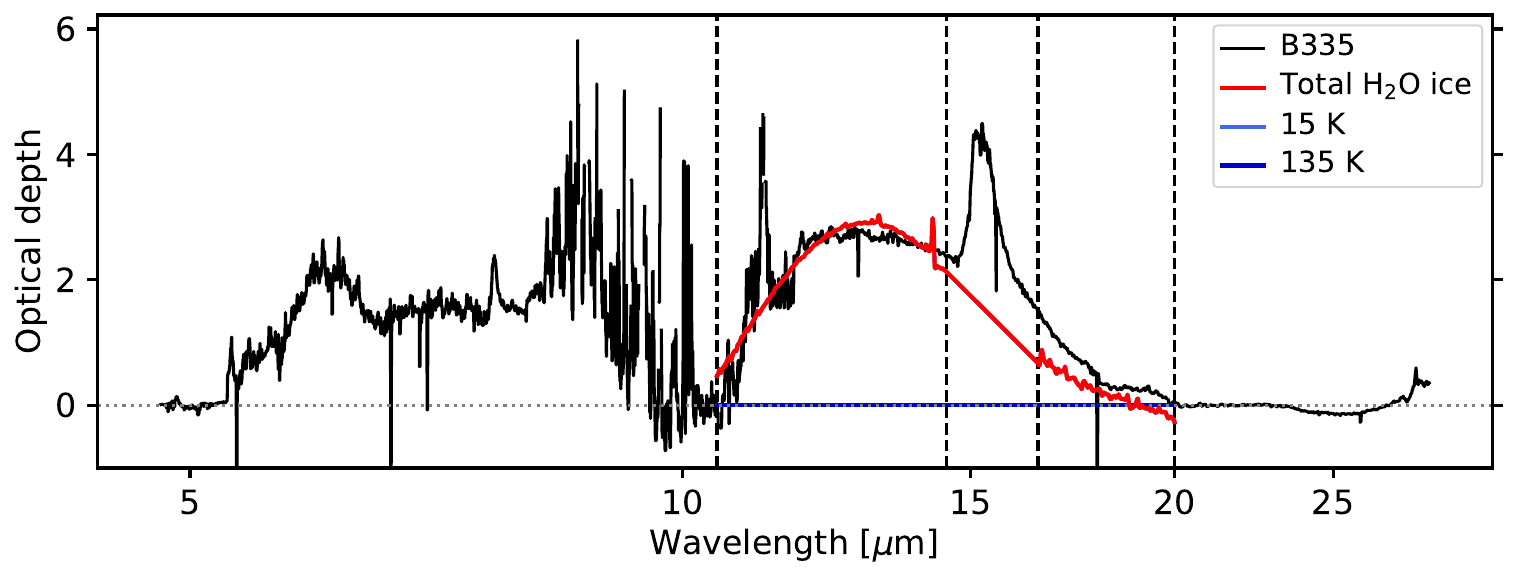}
    \includegraphics[width=0.48\linewidth]{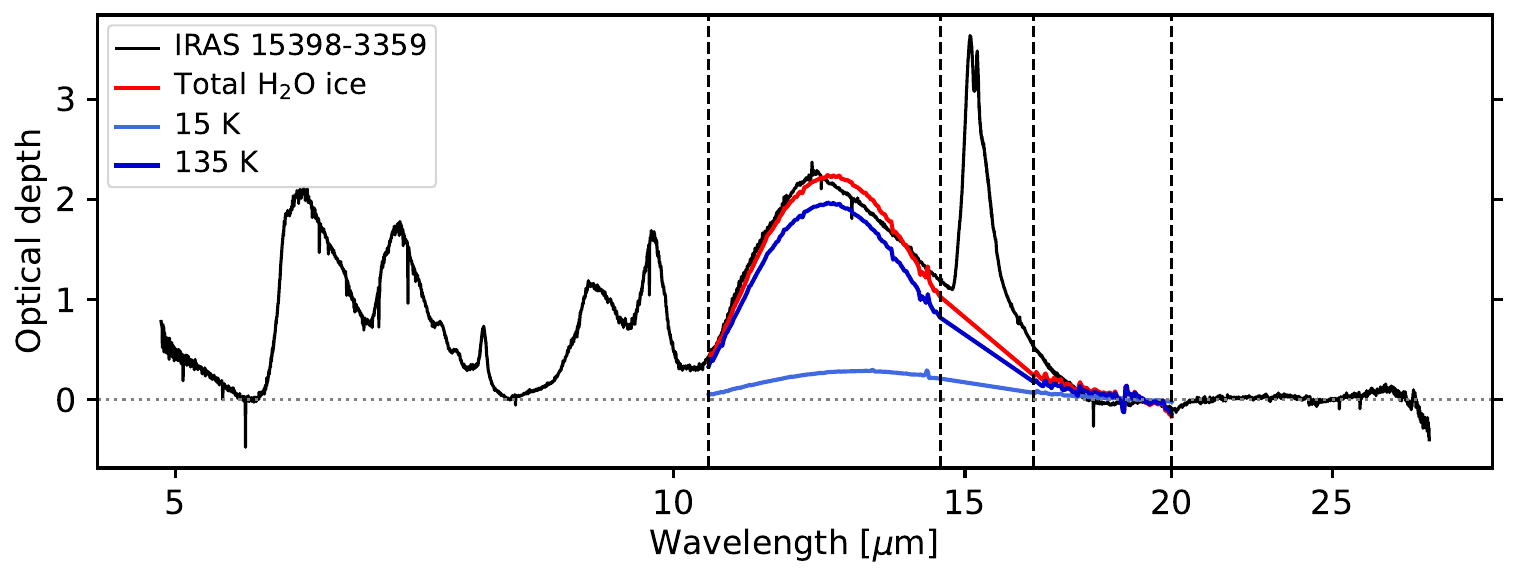}
    \includegraphics[width=0.48\linewidth]{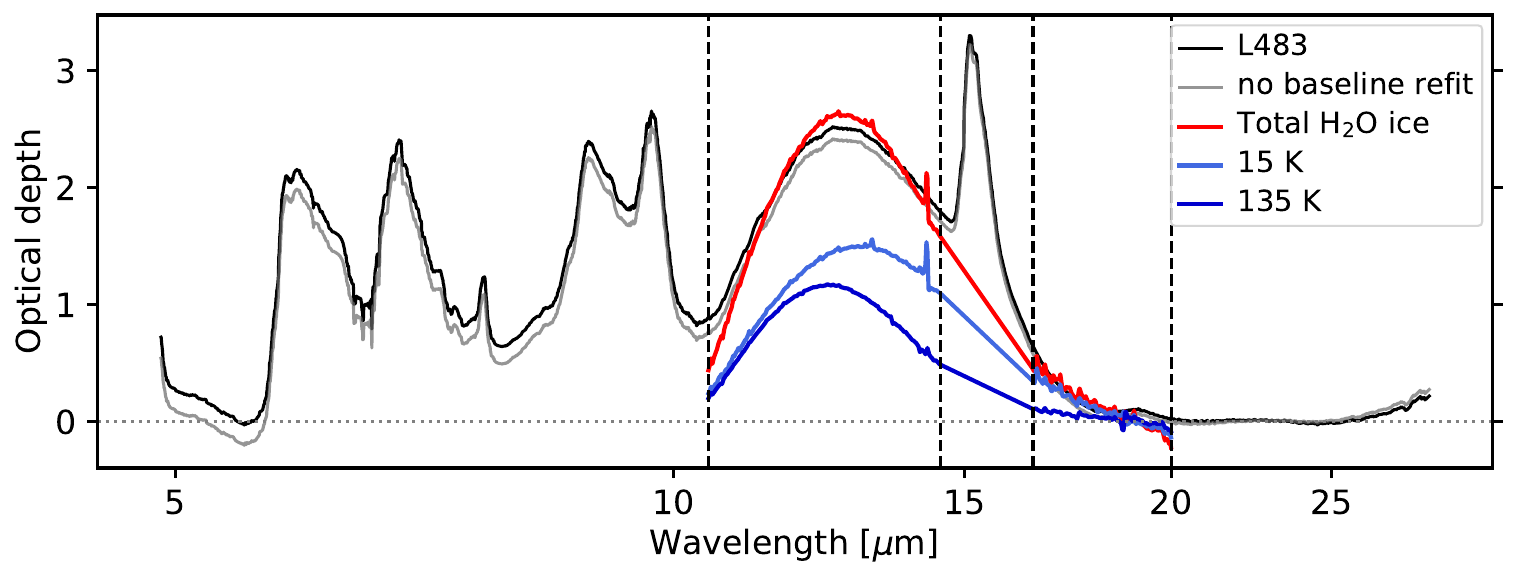}
    \includegraphics[width=0.48\linewidth]{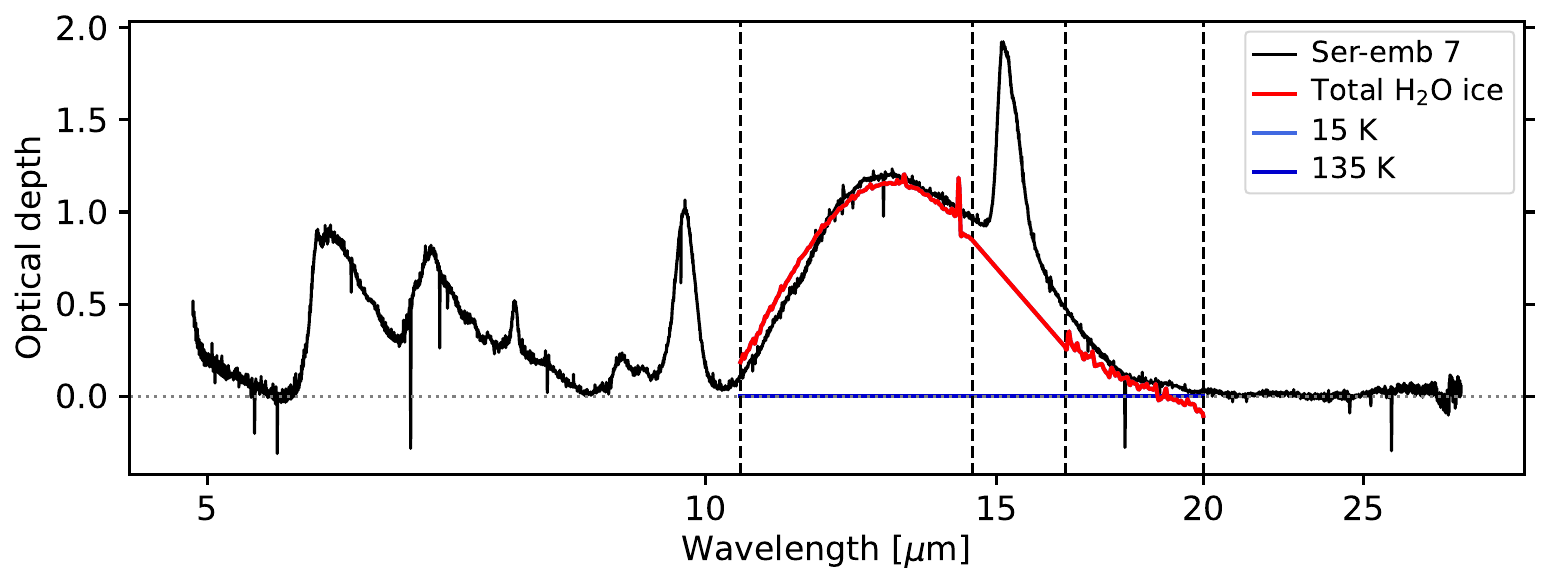}
    \caption{Best-fitting \water\ ice spectra in the CORINOS sources, shown in red.  The black line shows the silicate-subtracted optical depth spectra, while the fitted 15 K and 135 K \water\ ice components are shown in light blue and dark blue lines, respectively.  The light gray line in the L483 spectrum is the optical depth spectra without additional baseline calibration (see text).  The vertical dashed lines indicate the wavelength ranges, 10.5--14.5 \micron\ and 16.5--20 \micron, where the \water\ ice spectra are optimized.}
    \label{water_ice}
\end{figure*}






\end{document}